\documentstyle[pra,aps,epsfig,amstex]{revtex}
\begin{document}
\draft
\title{Geometrical Realization of Beutler-Fano formulas appearing in
eigenphase shifts and time delays in multichannel scattering}

\author{Chun-Woo Lee}
\address{Department of Chemistry, Ajou University, 5 Wonchun Dong,
	Suwon 442-749, KOREA}
\maketitle

\begin{abstract}

Recently, we showed that eigenphase shifts and
eigentime delays near a resonance for a system of one discrete state
and two continua are 
functionals of the Beutler-Fano formula using appropriate
dimensionless energy units and line profile indices and identified
parameters responsible for the avoided crossing of eigenphase shifts
and eigentime delays and also parameters responsible for the
eigentime delays due to a change in frame transformation.

In this paper, the geometrical realization of the Beutler-Fano formulas
is considered in the real three-dimensional Liouville space spanned by
the Pauli matrices $\sigma_x$, $\sigma_y$, $\sigma_z$ which are
orthogonal in the sense that ${\rm Tr} (\sigma_i \sigma_j ) = 2
\delta_{ij}$, where dynamic operators are vectors. Vectors
corresponding to the background $S^0$ matrix, $S$ matrix, and the time
delay matrix $Q$ form a spherical triangle whose vertex and edge
angles are  parameters pertaining to the frame
transformations among eigenchannels of $S^0$, $S$, $Q$ and eigenphase
shifts of $S^0$ and $S$ and the phase shift due to a resonance
scattering. The cotangent laws of the spherical triangle yield
Beutler-Fano resonance formulas appearing in eigenphase shifts and
time delays. Duality holding for the spherical triangle explains the
symmetry observed in the relations among parameters and provides a
systematic way of defining conjugate dynamic parameters. The spherical
triangle also shows the  rule of combining the 
channel-channel couplings in the background scattering with the
resonant interaction to give the avoided crossing interactions in the
curves of eigenphase shifts as functions of energy.

The theory developed in the previous and present papers is applied to
the vibrational predissociation of triatomic van der Waals molecules.

\end{abstract}

\section{INTRODUCTION}

Eigenphase shifts $\delta_i$ ($i$ = 1,2,$\dots , n$) of the $S$
matrix, defined as 
\begin{equation}
S = U e^{2i \delta} \tilde{U}
\label{S_out_bc}
\end{equation}
have been utilized as a tool for analyzing the
resonances\cite{Burke69,Truhlar87}. Eigenphase shifts and the
corresponding eigenchannels are also extensively used in various forms
in multichannel quantum-defect theory (MQDT) which is one of the most
powerful theories of resonance\cite{FanoBook}. In MQDT, the first
derivatives of eigenphase shifts as functions of energy are used for
various purposes\cite{Fano70}. The Lu-Fano plot is essentially a
plot of the energy derivative of an eigenphase
shift\cite{FanoBook}. The first derivative of an eigenphase shift as a
function of energy, called ``a partial delay time'', has also studied
as a relevant quantity to time delays in various
fields\cite{Fyodorov97}.

In spite of their wide use, studies of the behaviors of eigenphase
shifts and their energy derivatives in the neighborhood of a resonance
have not been done extensively, comparing to the studies of
photo-fragmentation cross sections and $S$ matrix itself. Eigenphase
shifts in the multichannel system are known to show complicated
behaviors near a resonance due to the avoided crossing between curves
of eigenphase shifts along the energy\cite{Burke69,Macek70}. In the
previous work, detailed studies of the behaviors of eigenphase shifts
and times delayed by collision were done for the system of one
discrete state and two continua\cite{Lee98}. Comparing to the system
of one discrete state and one continuum, the newly added one more
continuum brings about the new phenomena of the avoided crossing
between curves of eigenphase shifts and eigentimes delayed and of
times delayed due to the change in frame transformation along the
energy besides the resonance behavior. Thus the system of one discrete
state and two continua provides a prototypical system for the study of
two effects on the resonance phenomena, avoid crossings between
eigenphase shifts and eigentimes delayed as one effect and the change
in frame transformation as another one.

Previous work showed that eigenphase shifts and eigentimes delayed due
to the avoided crossing interaction and eigentimes delayed due to the
change in frame transformation are functionals of the Beutler-Fano
formula using the appropriate dimensionless energy units and line
profile indices\cite{Lee98}. Parameters representing the avoided
crossing of eigenphase shifts and eigentime delays were
identified. Parameters representing the eigentime delays due to a
change in frame transformation were shown to be described by the same
parameters. With the help of new parameters, the rotation angle
$\theta$ and rotation axis $\hat{n}$ for the $S$ matrix $\exp [
i(a+\theta \sigma \cdot \hat{n} )]$ were identified.  The time delay
matrix $Q$ was shown to be given as $Q = \frac{1}{2} \tau_r ({\bf 1} +
{\bf P}_a \cdot \boldsymbol{\sigma} + {\bf P}_f \cdot
\boldsymbol{\sigma} )$ where the first term is the time delay due to
the resonance, the second term is the one due to the avoided crossing
interaction and the last term is the one due to the change in frame
transformation.

Though previous work found that behaviors of eigenphase shifts and
eigentime delays as functions of energy  follow
Beutler-Fano's formula, it could not explain why they follow
Beutler-Fano's formula. Since the system considered in the previous
work is essentially two channel system (with one more discrete state),
an analogy with the spin system was made and utilized but not fully
exploited. One of the main purpose of the present paper is to exploit
the analogy further. Especially, the homomorphism of the spin model
with the three-dimensional rotation will be fully exploited to
construct geometric structures made up of dynamical parameters for
eigenphase shifts and eigentime delays and thus to derive the
Beutler-Fano's formula geometrically.  The geometrical realization
clarifies the ambiguities in relations and unexplained meanings of
dynamic parameters at the previous work since the geometric
constructions appeal to our intuition and are thus easy to understand
and provides means of viewing complicated relations in a simple
way. This clarification of  complicated dynamic relations through
the geometrical realization is another main goal of this paper.

Section II summarizes the previous results. Section III modifies the
previous results a little suitable for the geometrical
realization. Section IV gives a geometrical realization of the
previous results. Section V connects the geometrical relations with
the dynamical ones.  Section VI applies the theory developed in
Ref. \cite{Lee98} and the present paper to the vibrational
predissociation of triatomic van der Waals molecules. Section VII
summarizes and discusses the results.

\section{Summary of the previous result}

Ref. \cite{Lee98} examined eigenphase shifts and eigentime delays in
the neighborhood of an isolated resonance for the system of one discrete
state and two continua as a prototypical system for the study of the
combined effects of the resonance and the indirect continuum-continuum
interaction via a discrete state. The $S$ matrix for an isolated
resonance system is well known and, if the background $S^0$ matrix is
described by its eigenphase shifts $\delta^0$ as $S^0$= $U^0 e^{-2i
\delta^0} \tilde{U}^0$, takes the form\cite{TaylorBook}
\begin{equation}
S_{jk} = \sum_{l,m} U_{jl}^0 e^{-i \delta_l^0} \left( \delta_{lm} + i
 \frac{\sqrt{\Gamma_l \Gamma_m}}{E-E_0 -i \Gamma /2} \right) e^{-i
 \delta_m^0} \tilde{U}_{mk}^0 ,
\label{S_res}
\end{equation}
where $\Gamma_l$, $\Gamma$,$E_0$ are the partial decay width of a
resonance state into the $l$th background
eigenchannel\cite{ChannelTerm}, the total decay width $\sum_l
\Gamma_l$, and the resonance energy, respectively. Eq. (\ref{S_res})
is for the incoming wave boundary condition. The formula for the
outgoing wave boundary condition differs from Eq. (\ref{S_res}) in
that $i$ is replaced by $-i$.

By diagonalizing Eq. (\ref{S_res}), eigenphase shifts $\delta$ of
the $S$ matrix (=$U e^{-2i \delta }\tilde{U}$) for the system of one
discrete state and two continua were obtained as
\begin{equation}
2 \delta_{\pm} (E) = \sum_i \delta_i^{0} + \delta_r (E) \pm
\delta_a (E) ,
\label{eigenphase_shifts}
\end{equation}
where $\delta_r (E)$ is the well-known resonance phase shift due to
the modification of the scattering wave by the quasi-bound state and
given by $- \arctan (1/\epsilon_r )$ and $\delta_a (E)$ is the one due
to the modification of the scattering wave by the other wave through
the indirect interaction via the quasi-bound state and was found to be
given as a functional of the Beutler-Fano formula\cite{Fano61}
\begin{equation}
\cot \delta_a (E) =   - \cot \Delta_{12}^{0}  
\frac{\epsilon_a -q_{a} }{(1+\epsilon_a^2 )^{1/2}}  ,
\label{delta_a}
\end{equation}
in the dimensionless energy scale defined by
\begin{equation}
\epsilon_a \equiv  \frac{2(E-E_{a})}{\Gamma_{a}} , 
\label{e_a}
\end{equation}
where $\Gamma_a = 2 \sqrt{\Gamma_1 \Gamma_2}/ | \sin\Delta_{12}^{0} |$
and $E_a$ =$E_{0}+ \frac{\Delta\Gamma}{2}\cot\Delta_{12}^{0}$ 
($\Delta \Gamma = \Gamma_1 - \Gamma_2$, $\Delta_{12}^0 = \delta_1^0 -
\delta_2^0$).  Its form as a functional of the Beutler-Fano formula
can be shown more explicitly by using the Beutler-Fano function
\begin{equation}
f_{{\rm BF}} (\epsilon ,q) \equiv \frac{(\epsilon -q)^2}{1+\epsilon
^2} ,
\label{fBF}
\end{equation}
as 
\begin{equation}
\cot \delta_a (E ) = 
\left\{ \begin{array}{ll}
\cot \Delta_{12}^{0} \sqrt{f_{{\rm BF}} (\epsilon_a , q_{a} )}  &
  {\rm when}~\epsilon_a 
< q_a \\ 
- \cot \Delta_{12}^{0} \sqrt{f_{{\rm BF}}  (\epsilon_a , q_{a} )} 
& {\rm when}~\epsilon_a \ge q_a .
\end{array} \right.  
\label{delta_a_bf}
\end{equation}
The line profile index $q_a$ of the curve of $\delta_a (E)$ is given
by 
\begin{equation}
q_a = -\frac{ \Delta\Gamma} { 2\sqrt{\Gamma_1 \Gamma_2 } \cos
\Delta_{12}^{0} }.
\label{q_a}
\end{equation}
When $\epsilon_a$ = $q_a$, $\delta_a$ = $\pi /2$ and the
difference in abscissas of two eigenphase shift curves is $\pi /2$
which is the largest separation of two curves when $\delta_a$ is
defined up to $\pi$. Therefore, the line profile index $q_a$ also
stands for the energy of maximal avoidance of eigenphase shifts.
Eq. (\ref{S_res}) shows that the eigenphase sum $\delta_{\Sigma}$ is
given by $\delta_{\Sigma}$ = $\sum_i \delta_i^0$ + $\delta_r$ =
$\delta_{\Sigma}^0 $ + $\delta_r$, in conformity with Hazi's
formula\cite{Hazi79}.
 
Let us define ${\cal S}$ by $S$ = $U^0 {\cal S} \tilde{U}^0$.  Let
${\cal S}$ be diagonalized by the $V$ matrix composed of eigenvectors
corresponding to $\delta_+$ and $\delta_{-}$ as $V$ =
($v_{+},v_{-}$). The $V$ matrix was obtained as
\begin{equation}
V = \left( \begin{array}{cc} \cos\frac{\theta_{a}}{2} & - \sin
\frac{\theta_{a}}{2} \\ \sin \frac{\theta_{a}}{2} & \cos
\frac{\theta_{a}}{2} \end{array} \right) ,
\label{V}
\end{equation}
where $\theta_{a}$ is defined by
\begin{equation}
\cos\theta_{a} = -\frac{\epsilon_a}{\sqrt{1+\epsilon_a
    ^2}} , ~~
\sin\theta_{a} = \frac{1}{\sqrt{1+\epsilon_a
    ^2}} .
\end{equation}
Eigenvectors are independent of $q_{a}$. They depend only on
$\epsilon_a$. As $\epsilon_a$ varies from $-\infty$ through zero to
$\infty$, $\theta_{a}$ varies from zero through $\pi/2$ to $\pi$ and
$v_{+}$ varies from $\left( \begin{array}{cc}1\\ 0 \end{array}
\right)$ through $\frac{1}{\sqrt{2}} \left( \begin{array}{cc}1\\ 1
\end{array} \right)$ to $\left( \begin{array}{cc}0\\ 1 \end{array}
\right)$ .  Thus, at $\epsilon_a$ = 0 or at $E$ = $E_{0}$ +
$\frac{\Delta\Gamma}{2}\cot \Delta_{12}^{0}$, two background
eigenchannels are mixed equally. For this reason $\epsilon_a$ = 0 is
regarded as the avoided crossing point energy.  This energy does not
coincide with the energy $\epsilon_a$ = $q_a$ where two eigenphase
shift curves are separated most. Let $U$ = $U^0 V$, then $U$
diagonalizes the $S$ matrix, that is, the transform $\tilde{U} S U$ is
the diagonal matrix $e^{-2i \delta}$. The $U$ matrix is obtained from
the $V$ matrix by replacing $\theta_a$ with $\theta_a' = \theta_a +
\theta^{0}$, where $\theta^0$ parametrizes $U^{0}$ matrix as
\begin{equation}
 U^{0} = \left( \begin{array}{cc} \cos\frac{\theta^{0}}{2} & - \sin
\frac{\theta^{0}}{2} \\ \sin \frac{\theta^{0}}{2} & \cos
\frac{\theta^{0}}{2} \end{array} \right) ,
\label{U_0} 
\end{equation}

With the new parameters and Pauli's spin matrices, the $S$ matrix was
found to be expressible as
\begin{equation}
S = e^{-i\left(  \delta_{\Sigma} {\bf 1} +\delta_a 
\boldsymbol{\sigma} \cdot \hat{n}_a'  \right) } ,
\label{S_final}
\end{equation} 
where 
\begin{equation}
\hat{n}_a' = \hat{z} \cos \theta_a' + \hat{x} \sin
\theta_a' .
\label{na}
\end{equation}

Smith's time delay matrix $Q$ (=$i\hbar
S^{\dag}\frac{dS}{dE}$\cite{Smith,Qmatrix}) can be easily obtained by
substituting Eq. (\ref{S_final}) into its definition and was found to
consist of three terms
\begin{equation}
Q = \frac{1}{2} ({\bf 1} \tau_r + 
\boldsymbol{\sigma}\cdot \hat{n}_a' \tau_a 
+ \boldsymbol{\sigma}\cdot \hat{n}_f' \tau_f ) ,
\label{Q_af}
\end{equation}
one due to the resonance, one due to the avoided crossing interaction,
and one due to the change in frame transformation as a function of
energy, where
\begin{equation}
\hat{n}_f' = \hat{y} \times \hat{n}_a'
 \cos \delta_a  - \hat{y} \sin\delta_a  ,
\label{nf}
\end{equation}
and is orthogonal to $\hat{n}_a'$.  
 
The time delay due to the resonance takes a symmetric Lorentzian form
\begin{equation}
\tau_r (E ) = 2\hbar \frac{d \delta_r (E)}{dE} = \frac{4\hbar}
{\Gamma}\frac{1}{1+\epsilon_r ^2} ,
\label{tau_r}
\end{equation}
and the time delay due to the avoided crossing was found to take a
form of a functional of the Beutler-Fano formula
\begin{eqnarray}
\tau_a  (E ) &=& 
- \tau_r (E) \frac{\epsilon_r - q_{\tau}}
{\sqrt{( \epsilon_r - q_{\tau} )^2 + r^2 (1+\epsilon_r ^2 )}}
\label{tau_a}
\\
&=&
\left\{ \begin{array}{ll}
 \tau_r (E )
\sqrt{\frac{f_{{\rm BF}} (\epsilon_r ,q_{\tau})}{f_{{\rm BF}} (\epsilon_r
,q_{\tau} ) +r^2 }}
&{\rm when}~\epsilon_r \le q_{\tau}\\
- \tau_r (E )
\sqrt{\frac{f_{{\rm BF}} (\epsilon_r ,q_{\tau})}{f_{{\rm BF}} (\epsilon_r
,q_{\tau} ) +r^2 }}
&{\rm when}~\epsilon_r > q_{\tau} ,
\end{array} \right.  
\label{tau_a_bf}
\end{eqnarray}
where parameters $r$  and $q_{\tau}$ are defined by
\begin{equation}
r \equiv \frac{\sqrt{\Gamma ^2 - \Delta\Gamma ^2} }{\Delta \Gamma }
,\label{r_2} 
\end{equation}
\begin{equation}
q_{\tau} \equiv \frac{\Gamma}{\Delta\Gamma} \cot\Delta_{12}^{0} ,
\label{qtau} 
\end{equation}
The asymmetry of $\tau_a$ as a function of energy is brought about by
the nonzero value of $q_{\tau}$ which is proportional to the shift of
the avoided crossing point energy from the resonance one. Thus the
asymmetry of $\tau_a$ is caused by the mismatch in the positions of
the avoided crossing point and resonance energies.  The time delay due
to a change in frame transformation was found to take the following
form\cite{Convention1}
\begin{equation}
\tau_f (E) = \tau_r (E) \frac{|r|}{\sqrt{f_{\rm BF} (\epsilon_r
,q_{\tau} ) +r^2 }} .
\label{tauf_bf} 
\end{equation}

Because of the last term of
Eq. (\ref{Q_af}), eigenfunctions of the $Q$ matrix are different from
those of the $S$ matrix.  The eigentime delay sum, which is equal to
$\sum_i Q_{ii}$ = ${\rm Tr} Q$, is obtained as
\begin{equation}
\sum_i Q_{ii} = {\rm Tr} Q = \tau_r ,
\label{eigentime_delay_sum}
\end{equation}
since ${\rm Tr} \boldsymbol{\sigma}$ = 0.  The consideration of the
transforms $\tilde{U}^0 SU^0$ and $\tilde{U}^0 Q U^0$, which will be
denoted as ${\cal S}$ and ${\cal Q}$, will be proved to be more
convenient later for the geometric consideration. The transforms
${\cal S}$ and ${\cal Q}$ are the scattering and time delay matrices
with the background eigenchannel wavefunctions as a basis instead of
the asymptotic channel wavefunctions\cite{ChannelTerm}, whose forms in
terms of the new parameters and Pauli's spin matrices are the same as
those of $S$ and $Q$ but with vectors $\hat{n}_{a}$ and $\hat{n}_{f}$
which are obtained from $\hat{n}_a'$ and $\hat{n}_f'$ by replacing
$\theta_a'$ with $\theta_a$.

The connection of the time delay matrix $Q$ with the time delay
experienced by a wave packet was first considered by
Eisenbud\cite{Eisenbud} and extended by
others\cite{GoldbergerBook}. According to their work, $Q_{ii}$ is the
average time delay experienced by a wave packet injected in the $i$th
channel. Here, the average time delays due to the avoided crossing
interaction are given by $\tau_a \cos \theta_a' /2$ and $- \tau_a \cos
\theta_a' /2$. Similarly, the average time delays due to the change in
frame transformation are $- \tau_f \sin \theta_a' \cos \delta_a /2$
and $\tau_f \sin \theta_a' \cos \delta_a /2$. Time delays due to the
avoided crossing interaction and the change in frame transformation
are out of phase by $\pi /2$. Overall, $Q_{11}$ = $\frac{1}{2}
(\tau_r$ + $\tau_a \cos \theta_a'$ $-$ $\tau_f \sin \theta_a' \cos
\delta_a )$ and   $Q_{22}$ = $\frac{1}{2}
(\tau_r$ - $\tau_a \cos \theta_a'$ $+$ $\tau_f \sin \theta_a' \cos
\delta_a )$.

In analogy with the spin $\frac{1}{2}$ system, the time delay matrix
$Q$ was expressed in terms of polarization vectors and the Pauli
spin matrices as
\begin{equation}
Q = \frac{1}{2} \tau_r \left( {\bf 1} + {\bf P}_a \cdot
\boldsymbol{\sigma} + {\bf P}_f \cdot \boldsymbol{\sigma} \right) ,
\label{Qaf}
\end{equation}
where 
polarization vectors are defined by
\begin{equation}
{\bf P}_a = \frac{\tau_a}{\tau_r} \hat{n}_a' , ~~~ 
{\bf P}_f = \frac{\tau_f}{\tau_r} \hat{n}_f' .
\label{PaPf}
\end{equation}
Like the spin $\frac{1}{2}$ system, it was found that the absolute
values of ${\bf P}_a$ and ${\bf P}_f$ are restricted to $0 \le |{\bf
P}_a | \le 1$ and $0 \le | {\bf P}_f | \le 1$. In the present case a
complete depolarization means that eigentimes delays are the same
regardless of eigenchannels, while a complete polarization means that
eigentime delays are 0 for one eigenchannel and $\tau_r (E )$ for
another eigenchannel.
Eigenvectors for eigentime delays due to an avoided crossing
interaction and due to a change in frame transformation are orthogonal
to each other and contribute to the total eigentime delays as
$\sqrt{\tau_a ^2 + \tau_f ^2} = \tau_r \sqrt{|{\bf P}_a |^2 + |{\bf
P}_f |^2}$. It was found that
\begin{equation}
|{\bf P}_a |^2 + |{\bf P}_f |^2 =1 .
\label{Pt_magnitude}
\end{equation}
Since ${\bf P}_a$ and ${\bf P}_f$ are mutually orthogonal and $| {\bf
P} _a|^2 + |{\bf P} _f|^2 =1$, we can define a vector ${\bf P} _t =
{\bf P} _a + {\bf P} _f$, whose magnitude is unity.  Its formula may
be obtained straightforwardly but hardly used. Instead, the formula of
its transform
$\pmb{\cal P}_t$ = $\tilde{U}^0 {\bf P}_t U^0$ 
is exclusively used, which is much simpler and given as
\begin{equation}
\pmb{\cal P}_t \equiv \hat{n}_t = \left( \cos \Delta_{12}^0
\frac{\sqrt{\Gamma^2
- \Delta \Gamma^2}}{\Gamma}, - \sin \Delta_{12}^0
\frac{\sqrt{\Gamma^2 - 
\Delta \Gamma^2}}{\Gamma} , \frac{\Delta \Gamma}{\Gamma}\right) .
\label{nt}
\end{equation}
The transform $\pmb{\cal P}_t$ is the total polarization vector
for the time delay matrix ${\cal Q}$ (=$\tilde{U}^0 Q U^0$)
with background eigenchannels used as the basis. The similar
transforms $\pmb{\cal P}_a$ and $\pmb{\cal P}_f$ of ${\bf P}_a$ and
${\bf P}_f$ will be used later and satisfy the same relations
$\pmb{\cal P}_t$ = $\pmb{\cal P}_a$ + $\pmb{\cal P}_f$ and 
\begin{equation}
|\pmb{\cal P}_a |^2 + |\pmb{\cal P}_f |^2 =1 .
\label{Pt_cal_magnitude}
\end{equation}

With the total polarization vector, the time
delay matrix becomes
\begin{equation}
Q = \frac{1}{2} \tau_r ( {\bf 1} + {\bf P} _t \cdot {\bf
\sigma } ) .
\label{Q_t}
\end{equation}
Since $( {\bf P}_t \cdot \boldsymbol{\sigma})^2$ = 1, eigenvalues of
$Q$ or total eigentime delays are obtained as zero and $\tau_r$, the
time delayed by the resonance state. Though time delays due to an
avoided crossing interaction and a change in frame transformation are
asymmetric with respect to the resonance energy and therefore the
energies of the longest lifetimes are not matched with the resonance
energy, the energy of the longest overall eigentimes delayed is
exactly matched with the resonance energy.

\section{Preparation for the Geometrical Realization}
\label{Sec:Prep}

In the previous work, some of the interesting things were noticed but
could not be explained. Some of them are summarized below.
\begin{itemize}
\item Why are eigenvectors of the $S$ matrix independent of $q_a$ while
its eigenphase shifts are not?
\item Why do the energy behaviors of $\delta_a (E)$, $\tau_a (E)$,
and  $\tau_f (E)$
follow Beutler-Fano formulas?
\item Why does $\tau_a$ take the Beutler-Fano formula in the energy scale
of $\epsilon_r$ instead of $\epsilon_a$ in contrast to the case of
$\delta_a$ though the former is obtained as the derivative of the
latter. 
\item Why is $|{\bf P}_a |^2 + |{\bf P}_f |^2$ = 1 satisfied?
\item What is the meaning of the parameter $r^2$?
\end{itemize}
In the previous work, we got some help by making an analogy of the
system with a spin model, especially in interpreting the time delay
matrix $Q$ with polarization vectors ${\bf P}_a$ and ${\bf P}_f$ which
are borrowed from the spin model. But the analogy with the spin model
is not fully exploited. Here we show that by
exploiting the analogy further, we can give the explanations of the
above questions. In particular, we succeeded in giving the geometrical
realization of the Beutler-Fano formulas.
	
Before starting the geometrical realization of the previous results,
let us first rewrite some of the previous results suitable for the
geometrical realization. 

First, we notice that Eqs. (\ref{delta_a}) and (\ref{tau_a}) are
simpler than the corresponding Eqs. (\ref{delta_a_bf}) and
(\ref{tau_a_bf}). This indicates that the square root of the
Beutler-Fano formula (\ref{fBF}) seems to be more fundamental than the
original one. Next we notice that
Eq. ({\ref{delta_a})  resembles $\cot \delta_r$ = $-\epsilon_r$
and $\cot \theta_a$ = $-\epsilon_a$. Thus the square root of the
Beutler-Fano formula may be regarded as an energy parameter
$\epsilon_{\rm BF} (\epsilon , q, \theta^0 )$. Then
Eq. (\ref{delta_a}) takes the suggestive form
\begin{equation}
\cot \delta_a = - \epsilon_{\rm BF} (\epsilon_a , q_a ,
\Delta_{12}^0 ),
\end{equation}
where 
\begin{equation}
\epsilon_{\rm BF} (\epsilon_a , q_a ,
\Delta_{12}^0 ) = \cot \Delta_{12}^0 \frac{\epsilon_a - q_a
}{\sqrt{\epsilon_a^2 +1}}
\end{equation}
($\Delta_{12}^0$ is the value of $\delta_a$ at $\epsilon_a$
$\rightarrow$ $-\infty$). But there is a one drawback when the square
root of the Beutler-Fano formula is considered as an energy
parameter. It is not a monotonically increasing function of energy. It
has a minimum when $q>0$ and a maximum when $q<0$. Hence $\epsilon_{\rm
BF}$ will be considered here merely as a convenient notation.

Eq. (\ref{Pt_cal_magnitude}) suggests another angle $\theta_f$
satisfying $\pmb{\cal P}_a$ = $\hat{n}_a \cos \theta_f$ and $\pmb{\cal
P}_f$ = $\hat{n}_f \sin \theta_f$. Its cotangent is obtained as
\begin{equation}
\cot \theta_f  = - \frac{1}{r} \frac{\epsilon_r -
q_{\tau} } {\sqrt{\epsilon_r ^2 +1}} .
\label{cotf'}
\end{equation}
Eq. (\ref{cotf'}) indicates that $1/r$ becomes $\cot \theta_f$ at
$\epsilon_r$ $\rightarrow$ $- \infty$. The angle of $\theta_f$ at
$\epsilon_r$ $\rightarrow$ $- \infty$ is identified with the angle
which the polarization vector $\pmb{\cal P}_t$ or $\hat{n}_t$ makes
with $\hat{n}_a$. That angle will be denoted as
$\theta_t$. Eq. (\ref{nt}) shows that the angle $\theta_t$ is obtained
as
\begin{eqnarray}
\cos \theta_t &=& \frac{\Delta \Gamma}{\Gamma} \nonumber , \\ \sin
\theta_t &=& \frac{\sqrt{\Gamma ^2 - \Delta \Gamma ^2}}{\Gamma} 
\label{theta_t}
\end{eqnarray} 
and with it the spherical polar coordinate of $\hat{n}_t$ is given by
(1,$\theta_t$,$-\Delta_{12}^0$). Now with $\theta_t$,
Eq. (\ref{cotf'}) becomes
\begin{equation}
\cot \theta_f = - \cot \theta_t \frac{\epsilon_r - q_{\tau} }
{\sqrt{\epsilon_r ^2 +1}} = - \epsilon_{\rm BF} (\epsilon_r, q_{\tau},
\theta_t ) .
\label{cotf}
\end{equation}
With the new angle $\theta_f$, 
$\tau_a$ becomes $\tau_r \cos \theta_f$, which explains the
complicated form of $\tau_a$ as a functional of the Beutler-Fano
function in contrast to that of $\cot \delta_a$.

As a result of rewriting, we obtain four equations 
\begin{eqnarray}
\cot \delta_r &=& - \epsilon_r ,   \nonumber
\\
\cot \theta_a &=& - \epsilon_a ,
\nonumber
 \\
\cot \delta_a &=& - \epsilon_{\rm BF} (\epsilon_a , q_a ,
\Delta_{12}^0 ) ,
\nonumber  \\
\cot \theta_f &=& - \epsilon_{\rm BF} (\epsilon_r , q_{\tau} ,
\theta_t )  .
\label{cot_energy}
\end{eqnarray}
The use of the geometrical parameters, $\delta_r$ and $\theta_a$, in
place of $\epsilon_r$ and $\epsilon_a$ makes the geometrical
realization of dynamic relations possible. Our aim is to obtain the
dynamic formulas for $\epsilon_{\rm BF} (\epsilon_a , q_a ,
\Delta_{12}^0 )$ and $\epsilon_{\rm BF} (\epsilon_r , q_{\tau} ,
\theta_t ) $ by converting the geometric relations containing $\delta_a$
and $\theta_f$ back into dynamic
ones.  We will sometimes abbreviate $\epsilon_{\rm BF} (\epsilon_a
,q_a, \Delta_{12}^0 )$ as $\epsilon_{\rm BF,a}$ and $\epsilon_{\rm BF}
(\epsilon_r , q_{\tau} , \theta_t )$ as $\epsilon_{\rm BF,r}$. Before
ending this section, let us note the following formulas for the line
profile indices $q_a$ and $q_{\tau}$
\begin{eqnarray}
q_a &=& \frac{\cot \delta_a (\epsilon_a =0 )}{\cot \delta_a
(\epsilon_a \rightarrow -\infty )} , \nonumber\\
q_{\tau} &=& \frac{\cot \theta_f (\epsilon_r =0 )}{\cot \theta_f
(\epsilon_r \rightarrow -\infty )}  . 
\end{eqnarray}
They can also be expressed as  
\begin{eqnarray}
q_a &=& \frac{\cot \delta_a}{\cot \Delta_{12}^0} ~~~~{\rm when}~
  \theta_a = \frac{\pi}{2} ,
  \nonumber\\
q_{\tau} &=& \frac{\cot \theta_f}{\cot \theta_t} ~~~~{\rm when}~
  \delta_r = \frac{\pi}{2} .
\label{qaf_new_def}
\end{eqnarray}

\section{Geometrical Realization}

The geometrical realization is based on that to each unimodular
unitary matrix in the complex two-dimensional space, there is an
associated real orthogonal matrix representing a rotation in real
three-dimensional space\cite{Tinkham}. The general two-dimensional
unimodular unitary matrix can be written as $e^{- i \frac{\theta}{2}
\boldsymbol{\sigma} \cdot \hat{n}}$ as its determinant can be easily
shown to be unity using ${\rm Tr}(\boldsymbol{\sigma})$ = 0 and thus
unimodular by the definition of unimodularity. The associated real
orthogonal matrix will be denoted as $R_{\hat{n}} (\theta )$, the
rotation matrix about the vector $\hat{n}$ by an angle $\theta$
defined in an active sense.

Let us first consider the $S$ matrix. It is unitary but not unimodular
[det$(S)\ne 1$] and can not be associated with the pure rotation
alone. But after extracting det($S$) which is equal to $e^{-i
\delta_{\Sigma} }$ for isolated resonances (a similar formula holds
for overlapping resonances, where $\delta_r$ is replaced by the sum
over ones from all resonances\cite{Simonius74}), the remaining part of
the scattering matrix will be unimodular and may be associated with a
pure rotation. According to Eq. (\ref{S_final}), the remaining part is
$e^{-i \delta_a \boldsymbol{\sigma} \cdot \hat{n}_a}$ and may be
associated with the rotation about the vector $\hat{n}_a$ by an angle
2$\delta_a$. We will explore the possibility of this explanation of
the $S$ matrix in below by deriving Eq. (\ref{S_final}) in a more
systematic way.

In the previous section, the $S$ matrix was diagonalized by two
unitary matrices $U^0$ and $V$ given by Eqs. (\ref{U_0}) and
(\ref{V}). Actually the theorem in Ref. \cite{FanoRacahBook} limits
that $U^0$ and $V$ are real orthogonal since the $S$ matrix is
symmetric. With them, it is rewritten as
\begin{equation}
S = U^0 V e^{-2i \delta} \tilde{V} \tilde{U}^0 .
\label{S_U0V}
\end{equation}
Note that the two unitary transformations $U^0$ and $V$ can be written
in terms of Pauli spin matrices as
\begin{eqnarray}
U^0 &=& e^{-i \frac{\theta^0}{2} \boldsymbol{\sigma} \cdot \hat{y}}
,\nonumber \\ V &=& e^{-i \frac{\theta_a}{2} \boldsymbol{\sigma} \cdot
\hat{y}} .
\label{U0V}
\end{eqnarray}
Notice that argument matrices of two exponential functions are commute
and therefore $U^0V$ = $e^{-i \frac{1}{2} (\theta_a + \theta^0)
\boldsymbol{ \sigma} \cdot \hat{y}}$. As before, let us denote
$\theta_a + \theta^0$ as $\theta_a'$.  The diagonalized matrix $e^{-2i
\delta}$ of the $S$ matrix can be expressed in terms of Pauli matrices
as
\begin{equation}
e^{-2i \delta} = \left( \begin{array}{cc} e^{-2i \delta_{+}} & 0 \\ 0
 & e^{-2i \delta_{-}} \end{array} \right) = e^{-i(\delta_{\Sigma}{\bf
 1} + \delta_a \boldsymbol{\sigma } \cdot \hat{z}) } .
\label{Sdiag}
\end{equation}
Substituting Eqs. (\ref{U0V}) and (\ref{Sdiag}) into
Eq. (\ref{S_U0V}), we obtain
\begin{equation}
S = e^{-i\frac{\theta_a'}{2} \boldsymbol{\sigma } \cdot \hat{y}}
e^{-i(\delta_{\Sigma}{\bf 1} + \delta_a \boldsymbol{\sigma } \cdot
\hat{z}) } e^{i \frac{\theta_a'}{2} \boldsymbol{\sigma } \cdot
\hat{y}}.
\label{S}
\end{equation}

In order to give the geometrical interpretation to Eq. (\ref{S}), a
long preliminary exposition is necessary. Let us start with
considering $\boldsymbol{\sigma} \cdot {\bf r}$ and transform it into
a new matrix $\boldsymbol{\sigma} \cdot {\bf r}'$ by a general 2
$\times$ 2 unitary transformation $e^{-i \frac{\theta}{2}
\boldsymbol{\sigma} \cdot \hat{n}}$ as follows
\begin{equation}
e^{-i \frac{\theta}{2} \boldsymbol{\sigma} \cdot \hat{n}} \,
\boldsymbol{\sigma} \cdot {\bf r}\, e^{i \frac{\theta}{2}
\boldsymbol{\sigma} \cdot \hat{n}} = {\bf \sigma}\cdot {\bf r}' .
\label{r}
\end{equation}
The left hand side of Eq. (\ref{r}) can be calculated
using\cite{MerzbacherBook}  
\begin{equation}
e^{i\hat{S}}
\hat{O} e^{-i\hat{S}} = \hat{O} + i [\hat{S},\hat{O}] +
\frac{i^2}{2!} [\hat{S},[\hat{S},\hat{O}]] +
\frac{i^3}{3!} [\hat{S},[\hat{S},[\hat{S},\hat{O}]]] + \dots
\label{exp_sim}
\end{equation}
and $(\boldsymbol{\sigma} \cdot {\bf a}) (\boldsymbol{\sigma} \cdot
{\bf b} )$ = ${\bf a} \cdot {\bf b}$ + $i \boldsymbol{\sigma} \cdot (
{\bf a} \times {\bf b} )$ and the result is that ${\bf r}'$ is just
the vector obtained from ${\bf r}$ by the three dimensional rotation
matrix $R_{\hat{n}} (\theta )$ about the vector $\hat{n}$ by $\theta$ in
an active sense as
\begin{equation}
{\bf r}' = R_{\hat{n}} (\theta ) {\bf r} .
\label{rp}
\label{rotation}
\end{equation}
Only in the form of the similarity transformation (\ref{r}) the
homomorphism that a 2 $\times$ 2 unimodular unitary matrix is
associated with a three dimensional rotation holds.  According to this
interpretation, the unitary transformations $U^0$ and $V$ for the
symmetric $S$ matrix (\ref{S_U0V}) correspond to the three-dimensional
rotations about the $y$ axis through angles $\theta^0$ and $\theta_a$,
respectively, and their overall effect $U^0V$ is equal to the rotation
about the $y$ axis by $\theta_a'$ = $\theta_a + \theta^0$.  Therefore,
the original frame transformation for the symmetric $S$ matrix becomes
as a rotation about the $y$ axis in the ``hypothetical'' real
three-dimensional space. This hypothetical real three-dimensional
space is different from the Hilbert space and called the Liouville
space. It is the space spanned by the set of vectors $\sigma_x$,
$\sigma_y$, and $\sigma_z$ which are orthogonal in the sense that
\begin{equation}
{\rm Tr} (\sigma_i \sigma_j ) = 2 \delta_{ij} ,
\end{equation}
and extensively studied in Ref. \cite{Fano57}. Any traceless 2
$\times$ 2 Hermitian matrices, for example $h$, can be expanded in
this vector space as $h$ = $x \sigma_x + y \sigma _y + z \sigma_z$ =
($x,y,z$) = $\boldsymbol{\sigma} \cdot {\bf r}$.  We can lift the
restriction of traceless in Hermitian matrices if we include the unit
matrix ${\bf 1}$ as another basic vector in addition to $\sigma_x$,
$\sigma_y$, $\sigma_z$. Then the three-dimensional Liouville space is
a subspace of this four-dimensional Liouville space. Note that two
subspace \{${\bf 1}$\} and \{$\sigma_x , \sigma_y , \sigma_z $\} are
orthogonal and therefore either the subspace \{$\sigma_x , \sigma_y ,
\sigma_z$\} or \{${\bf 1 }$\}, or the whole space may be chosen freely
depending on the situation without having any trouble.

Now, Eq. (\ref{r})
can be viewed in two ways. It can be viewed as a rotation of the
vector ${\bf r}$ into ${\bf r}'$ by the rotation matrix $R_{\hat{n}}
(\theta )$ as expressed in Eq. (\ref{rotation}). Or it can be viewed as
the transformation from the $xyz$ coordinate system to the $x'y'z'$
coordinate system by the rotation matrix $R_{\hat{n}} ( - \theta
)$. The latter view, though obvious, can be shown to be true using the
following mathematical transformation. Let us regard
$\boldsymbol{\sigma}$ and ${\bf r}$ as the column vectors. Then the
scalar product $\boldsymbol{\sigma} \cdot {\bf r}'$ can be written as
a matrix multiplication of the row vector $\tilde{
\boldsymbol{\sigma}}$ with the column vector ${\bf r}'$, namely, $
\boldsymbol{\sigma} \cdot {\bf r}' = \tilde{\boldsymbol{\sigma}} {\bf
r}'$. The support for the view of the coordinate transformation is
obtained by the following transformation
\begin{equation}
\tilde{\boldsymbol{\sigma}} {\bf
r}' = \tilde{\boldsymbol{\sigma}} R_{\hat{n}} (\theta ){\bf r} =
\widetilde{ [ R_{\hat{n}} (- \theta ) \boldsymbol{\sigma} ]} {\bf r} .
\end{equation}
Since the diagonalization of the operator $\boldsymbol{\sigma} \cdot
{\bf r}$ yields its eigenchannels, the vector ${\bf r}$ in the
three-dimensional Liouville space is enough to
uniquely specify the eigenchannels of the traceless Hermitian
matrix $\boldsymbol{\sigma} \cdot {\bf r}$. Conversely, eigenchannels
may be regarded as a vector in the 
three-dimensional Liouville space. 

Since any real orthogonal frame transformation is of the form like
Eq. (\ref{V}), it may be generally said that a real orthogonal frame
transformation in the complex two-dimensional space corresponds to a
rotation about the $y$ axis in the real three-dimensional Liouville
space. Since the matrix corresponding to any 2 $\times$ 2 Hermitian
operator is diagonal in the basis of eigenchannels by definition of
eigenchannels and can be written as $a{\bf 1} + b \boldsymbol{\sigma}
\cdot \hat{z}$, a dynamical process along an eigenchannel corresponds
to a process along the $z$ axis in the real three-dimensional
Liouville space and leads to a variation in length of the vector.
Thus the $y$ axis in the real three-dimensional Liouville space can be
regarded as the axis for the real orthogonal frame transformations and
the $z$ axis as the axis for the dynamical processes along
eigenchannels.

We have theorem that $C e^{B} C^{-1}$ = $e^{CBC^{-1}}$ which can be
easily proved by using Eq. (\ref{exp_sim}). Using this theorem and
Eq. (\ref{r}), we have
\begin{equation}
e^{-i \frac{\theta}{2} \boldsymbol{\sigma} \cdot \hat{n}} e^{i
\boldsymbol{\sigma} \cdot {\bf r} } e^{i \frac{\theta}{2}
\boldsymbol{\sigma} \cdot \hat{n}} = e^{i \boldsymbol{\sigma} \cdot
{\bf r}' } .
\label{exp_r}
\end{equation}
Using Eq. (\ref{exp_r}) and Eq. (\ref{rp}), Eq. (\ref{S}) becomes
Eq. (\ref{S_final}) with $\hat{n}_a'$ now interpreted as 
\begin{equation}
\hat{n}_a' = R_{\hat{y}} (\theta_a' ) \hat{z} .
\label{na_rot}
\end{equation}
Or $\hat{n}_a'$ can be regarded as the $z'$ axis in the $x'y'z'$
coordinate system, i.e., $\hat{n}_a'$ = $\hat{z}'$. Let $S = e^{-2i
\boldsymbol{\Delta}'}$, i.e., $\boldsymbol{\Delta}'$ = $\frac{1}{2} (
\delta_{\Sigma} \, {\bf 1} + \delta_a \boldsymbol{\sigma} \cdot
\hat{n}_a' )$. $\boldsymbol{\Delta}'$ is a vector in the
four-dimensional Liouville space. Or, if we exclude the isotropic part
in $\boldsymbol{\Delta}'$, $\frac{1}{2} \delta_a \boldsymbol{\sigma}
\cdot \hat{n}_a'$ is a vector in the three-dimensional Liouville space
which may be obtained by rotating the vector $\frac{1}{2} \delta_a
\boldsymbol{\sigma} \cdot \hat{z}$ about the $y$ axis by an angle
$\theta_a'$.  The $\frac{1}{2} \delta_{\Sigma} \, {\bf 1}$ term in
$\boldsymbol{\Delta}'$ gives the phase shift owing to the isotropic
influence of the background potential scattering and the resonance on
eigenchannels.  Likewise, the $\frac{1}{2} \delta_a
\boldsymbol{\sigma} \cdot \hat{n}_a'$ term gives the phase shifts
owing to the anisotropic influence of the background scattering
potential and the resonance on eigenchannels. Therefore, the length
$\frac{1}{2} \delta_a$ of
the anisotropic term in the three-dimensional Liouville space
denotes the degree of anisotropic influence on eigenphase shifts by
the background scattering and the resonance.

Let us now consider the time delay matrix. If the time delay matrix is
written in Lippmann's suggestive form\cite{Lippman66}, $Q = S^{+} \tau
S$ ($\tau$ is the time operator defined by $i \hbar
\frac{\partial}{\partial E}$), it is apparent that the unitary matrix
which gives the similarity transformation is now the $S^+$ matrix and
can be associated with the rotation matrix according to the theorem
mentioned above when det($S$) is extracted from it. Using the relation
\begin{equation}
\frac{d^r}{dz^r} \left( e^{Az} \right) = A^r e^{Az} = e^{Az} A^r ,
\end{equation}
Eq. (\ref{S}) is easily differentiated with respect to energy to yield
\begin{equation}
\frac{dS}{dE} = -i \frac{d\delta_{\Sigma}}{dE} S -i
\frac{d\delta_a}{dE} e^{-i \frac{\theta_a'}{2} \boldsymbol{\sigma}
\cdot \hat{y}} e^{-i \delta_a \boldsymbol{\sigma} \cdot \hat{z}} {\bf
\sigma} \cdot \hat{z} e^{i \frac{\theta_a'}{2} \boldsymbol{\sigma}
\cdot \hat{y}} + \frac{i}{2} \frac{d \theta_a'}{dE} \left( S
\boldsymbol{\sigma} \cdot \hat{y} - \boldsymbol{\sigma} \cdot \hat{y}
S \right) .
\label{dSdE}
\end{equation}
By multiplying (\ref{dSdE}) with the adjoint of (\ref{S}),
the $Q$ matrix becomes 
\begin{equation}
Q = i\hbar S^+ \frac{dS}{dE} = \hbar \frac{d\delta_{\Sigma}}{dE} {\bf
1} +
\hbar \frac{d\delta_{a}}{dE} e^{-i \frac{\theta_a'}{2}
\boldsymbol{\sigma} \cdot \hat{y}} \boldsymbol{\sigma} \cdot \hat{z}
e^{i \frac{\theta_a'}{2} \boldsymbol{\sigma} \cdot \hat{y}} +
\frac{\hbar}{2} \frac{d\theta_a'}{dE} \left( S^+ \boldsymbol{\sigma}
\cdot \hat{y}S - \boldsymbol{\sigma} \cdot \hat{y} \right) ,
\label{Q_deriv1}
\end{equation}
where use is made of the fact that $S$ is unitary and thus $S^+ S$ = 1
for the first and third terms. The matrix multiplication in the second
term of the right hand side of Eq. (\ref{Q_deriv1}) is just ${\bf
\sigma} \cdot \hat{n}_a'$ as was already done before. The sum of the
first and second term is the time delay due to the energy derivatives
of the eigenphase shifts and called  ``the partial delay times'' by some
group of people\cite{Fyodorov97}. 

The parenthesized part of the third term is the time delay due to the
change in frame transformation and has an interference effect between
two contributions, one due to the change in frame transformation from
the asymptotic channels to the background eigenchannels $\langle
\psi_E^{-(k)} | \psi_E^{(l)} \rangle$ ($k,l$ = 1,2,...,n) and the
other due to the change in frame transformation from the background
eigenchannels to the asymptotic eigenchannels $\langle \psi_E^{(l)} |
\psi_E^{-(m)} \rangle$ ($l,m$ = 1,2,...,n). The change in frame
transformation does not take place in the direction of the rotation
axis given by $\hat{y}$ but in the rotation angle since the rotation
$y$ axis is fixed in the Liouville space as energy varies.  The first
contribution has the term $S^{+} \boldsymbol{\sigma} \cdot \hat{y} S$
which is the similarity transformation of the operator
$\boldsymbol{\sigma} \cdot \hat{y}$ by $S$. Substituting Eq. (\ref{S})
for $S$, this term becomes
\begin{equation}
S^{+} \boldsymbol{\sigma} \cdot \hat{y} S = e^{-i \frac{\theta_a'}{2}
\boldsymbol{\sigma} \cdot \hat{y}} e^{i \delta_a \boldsymbol{\sigma}
\cdot \hat{z}} \boldsymbol{\sigma} \cdot \hat{y} e^{-i \delta_a
\boldsymbol{\sigma} \cdot \hat{z}} e^{i \frac{\theta_a'}{2}
\boldsymbol{\sigma} \cdot \hat{y}} ,
\label{Q_deriv2}
\end{equation}
where use is made of that $e^{i \frac{\theta_a'}{2}
\boldsymbol{\sigma} \cdot \hat{y}}$ and $\boldsymbol{\sigma} \cdot
\hat{y}$ are commutative.  The scalar factor $e^{- i \delta_{\Sigma}}$
in $S$ does not appear in Eq. (\ref{Q_deriv2}) as it is multiplied by
its complex conjugation in $S^{+}$ to become unity.  According to the
theorem, the unitary transformations in the right hand side of
Eq. (\ref{Q_deriv2}) correspond to two consecutive rotations, at first
about the $z$ axis by $-2 \delta_a$ and then about the $y$
axis by $\theta_a$. By the first rotation, $\boldsymbol{\sigma} \cdot
\hat{y}$ becomes $\boldsymbol{\sigma} \cdot \left( \hat{x} \sin 2
\delta_a + \hat{y} \cos 2 \delta_a \right)$, i.e.,
\begin{equation}
e^{i\delta_a \boldsymbol{\sigma} \cdot \hat{z}} \boldsymbol{\sigma}
\cdot \hat{y} e^{-i \delta_a \boldsymbol{\sigma} \cdot \hat{z}} =
\boldsymbol{\sigma} \cdot \left[ R_{\hat{z}} (-2 \delta_a ) \hat{y}
\right] = \boldsymbol{\sigma} \cdot \left( \hat{x} \sin 2 \delta_a +
\hat{y} \cos 2 \delta_a \right) .
\label{rot_y_about_z}
\end{equation}
By substituting Eq. (\ref{rot_y_about_z}) into Eq. (\ref{Q_deriv2})
and $ \boldsymbol{\sigma} \cdot \hat{y} $'s being replaced with $e^{-i
\frac{\theta_a'}{2} \boldsymbol{\sigma} \cdot \hat{y}}
\boldsymbol{\sigma} \cdot \hat{y} e^{i \frac{\theta_a'}{2}
\boldsymbol{\sigma} \cdot \hat{y}}$,
\begin{eqnarray}
S^+ \boldsymbol{\sigma} \cdot \hat{y}S - \boldsymbol{\sigma} \cdot
\hat{y} &=& e^{-i \frac{\theta_a'}{2} {\bf \sigma} \cdot \hat{y}}
\left[ \boldsymbol{\sigma} \cdot \left( \hat{x} \sin 2 \delta_a +
\hat{y} \cos 2 \delta_a \right) - \boldsymbol{\sigma} \cdot \hat{y}
\right] e^{i \frac{\theta_a'}{2} {\bf \sigma} \cdot \hat{y}} \nonumber
\\ &=& e^{-i \frac{\theta_a'}{2} {\bf \sigma} \cdot \hat{y}} \left[ 2
\sin \delta_a \boldsymbol{\sigma} \cdot \left( \hat{x} \cos \delta_a -
\hat{y} \sin \delta_a \right) \right] e^{i \frac{\theta_a'}{2} {\bf
\sigma} \cdot \hat{y}} .
\label{Q_third}
\end{eqnarray}
The bracketed part of Eq. (\ref{Q_third}) is equal to the rotation of
the $x$ axis about the $z$ axis by $- \delta_a$ multiplied
by $2 \sin \delta_a$, which is the overall effect of the
interference. Fig. \ref{fig:frmchg} shows this process of interference
as a vector addition in the  three-dimensional Liouville space.
The time delay due to the change in frame transformation, the third
term of the right hand side of Eq. (\ref{Q_deriv1}), becomes
\begin{eqnarray}
&& \hbar \sin \delta_a \frac{d\theta_a'}{dE} 
e^{-i \frac{\theta_a'}{2} \boldsymbol{\sigma} \cdot \hat{y}}
\left( \hat{x} \cos \delta_a - \hat{y} \sin \delta_a \right)
e^{i \frac{\theta_a'}{2} \boldsymbol{\sigma} \cdot \hat{y}} \nonumber \\
&=& \hbar \sin \delta_a \frac{d\theta_a'}{dE} 
e^{-i \frac{\theta_a'}{2} \boldsymbol{\sigma} \cdot \hat{y}}
e^{i \frac{\delta_a}{2} \boldsymbol{\sigma} \cdot \hat{z}}
\boldsymbol{\sigma} \cdot \hat{x}
e^{-i \frac{\delta_a}{2} \boldsymbol{\sigma} \cdot \hat{z}}
e^{i \frac{\theta_a'}{2} \boldsymbol{\sigma} \cdot \hat{y}}
 \nonumber \\
&=& \hbar \sin \delta_a \frac{d\theta_a'}{dE} 
e^{i \frac{\delta_a}{2} \boldsymbol{\sigma} \cdot \hat{z'}}
\boldsymbol{\sigma} \cdot \hat{x}'
e^{-i \frac{\delta_a}{2} \boldsymbol{\sigma} \cdot \hat{z'}} \nonumber \\
&=&  \hbar \sin \delta_a \frac{d\theta_a'}{dE} \boldsymbol{\sigma} \cdot
\hat{x}'' ,
\label{Q_f2}
\end{eqnarray}
where the second equality is obtained by applying Eq. (\ref{r}) twice
to the matrix term to obtain $\boldsymbol{\sigma} \cdot
\left[R_{\hat{y}} (\theta_a' ) R_{\hat{z}} (- \delta_a ) \hat{x}
\right]$ which becomes $\boldsymbol{\sigma} \cdot R_{\hat{z}'}
(-\delta_a ) \hat{x}'$ and then by applying Eq. (\ref{r}) again.  In
the last equality of Eq. (\ref{Q_f2}), we introduced another new $x''y''z''$
coordinate system which is obtained from the $x'y'z'$ coordinate
system by the rotation $R_{\hat{z}'} (\delta_a )$ in the passive sense.
In the active sense, $\hat{x}''$ = $R_{\hat{z}'} (- \delta_a )
\hat{x}'$.

Substituting
Eq. (\ref{Q_f2}) into Eq. (\ref{Q_deriv1}), the time delay matrix $Q$
is obtained as
\begin{equation}
Q = \hbar \left( {\bf 1} \frac{d \delta_{\Sigma}}{dE} +
\boldsymbol{\sigma } \cdot \hat{z}' \frac{d \delta_a} {dE} +
\boldsymbol{\sigma } \cdot \hat{x}'' \sin \delta_a \frac{d
\theta_a}{dE} \right) ,
\label{Q_final2}
\end{equation}
and is equal to Eq. (\ref{Q_af}) when $\hat{z}'$ and $\hat{x}''$ are
identified with the unit vectors $\hat{n}_a'$ and $\hat{n}_f'$
($\hat{n}_{\theta_a'}$ and $\hat{n}_{\theta_a'}^{\perp}$ in
Ref. \cite{Lee98}) and $\hbar d\delta_{\Sigma} /dE$ (= $\hbar
d\delta_r /dE$), $\hbar \boldsymbol{\sigma} \cdot \hat{z}' d\delta_a
/dE$, and $\hbar \sin \delta_a \boldsymbol{\sigma} \cdot \hat{x}''
d\theta_a /dE$ are identified with $\frac{1}{2} \tau_r $, $\frac{1}{2}
\tau_r {\bf P}_a$ and $\frac{1}{2} \tau_r {\bf P}_f$, respectively. By
substituting ${\bf P}_a$ = $\hat{z}' \cos \theta_f$ = $\hat{z}'' \cos
\theta_f$ and ${\bf P}_f$ = $\hat{x}'' \sin \theta_f$,
Eq. (\ref{Q_final2}) can also be transformed as follows
\begin{eqnarray}
Q &=& \frac{1}{2} \tau_r \left[ {\bf 1} + \boldsymbol{\sigma} \cdot
\left( \hat{z}'' \cos \theta_f + \hat{x}'' \sin \theta_f \right)
\right] \nonumber \\ &=& \frac{1}{2} \tau_r \left( {\bf 1} + e^{-i
\frac{\theta_f}{2} \boldsymbol{\sigma} \cdot \hat{y}''}
\boldsymbol{\sigma} \cdot \hat{z}'' e^{i \frac{\theta_f}{2}
\boldsymbol{\sigma} \cdot \hat{y}''} \right) \nonumber \\ &=&
\frac{1}{2} \tau_r \left( {\bf 1} + \boldsymbol{\sigma} \cdot
\hat{z}''' \right) ,
\label{Q_final3}
\end{eqnarray}
where still another $x'''y'''z'''$ coordinate system is introduced. In
the active sense, $\hat{z}'''$ = $R_{\hat{y}''} (\theta_f )
\hat{z}''$. 
Eqs. (\ref{Q_final2}) and (\ref{Q_final3}) tells us that time delay
matrices due to the avoided crossing interaction,due to the change in
frame transformations and the total time delay matrix take simplest
form in the $x'y'z'$, $x''y''z''$, and $x'''y'''z'''$ coordinate
systems, respectively.

Eq. (\ref{Q_final3}) equals Eq. (\ref{Q_t}) and therefore $\hat{z}'''$
equals ${\bf P}_t$. The vector $\hat{z}'''$, and accordingly ${\bf
P}_t$, can be obtained from $\hat{z}$
by successive rotations by
\begin{equation}
{\bf P}_t = \hat{z}''' = R_{\hat{y}''} (\theta_f ) R_{\hat{z}'} (-
\delta_a ) R_{\hat{y}} (\theta_a' ) \hat{z} = R_{\hat{y}''} (\theta_f
) R_{\hat{y}} (\theta_a' ) \hat{z} .
\label{nt_euler}
\end{equation}  
As mentioned before, it is better to consider $\pmb{\cal P}_t$ =
$\tilde{U}^0 {\bf P}_t U^0$ rather than ${\bf P}_t$ itself since the
formula for the former is simpler that that for the latter. $\pmb{\cal
P}_t$ is the polarization vector pertaining to ${\cal Q} = \tilde{U}^0
Q U^0$ which is the time delay matrix in the basis of background
eigenchannel wavefunctions. This suggests that it may be better to
take the background eigenchannel wavefunctions rather than the
asymptotic channel wavefunctions as a starting channel basis. From now
on, let us redefine the $xyz$ coordinate system as the coordinate
system pertaining to the background eigenchannels. Let us use the name
of $x^0 y^0 z^0$ coordinate system as that pertaining to the
asymptotic channels. Definitions of other coordinate systems remain
unchanged. With this redefinition of notation, the formulas
for ${\cal S}$ and ${\cal Q}$ corresponding to Eqs. (\ref{S_final})
and (\ref{Q_af}) becomes
\begin{eqnarray}
{\cal S} &=& e^{-i\frac{\theta_a}{2} \boldsymbol{\sigma } \cdot
\hat{y}} e^{-i(\delta_{\Sigma}{\bf 1} + \delta_a \boldsymbol{\sigma }
\cdot \hat{z}) } e^{i \frac{\theta_a}{2} \boldsymbol{\sigma } \cdot
\hat{y}} = e^{-i\left( \delta_{\Sigma}{\bf 1} +\delta_a
\boldsymbol{\sigma} \cdot \hat{n}_a \right) } , 
\label{S_new_basis}
\\ 
{\cal Q}
&=& \frac{1}{2} (\tau_r {\bf 1} + \tau_a \boldsymbol{\sigma}\cdot
\hat{n}_a + \tau_f \boldsymbol{\sigma}\cdot \hat{n}_f ) = 
\frac{1}{2}
\tau_r \left( {\bf 1} + \pmb{\cal P}_a \cdot \boldsymbol{\sigma} +
\pmb{\cal P}_f \cdot \boldsymbol{\sigma} \right) 
= \frac{1}{2}
\tau_r \left( {\bf 1} + \pmb{\cal P}_t \cdot \boldsymbol{\sigma} 
\right) ,
\label{Q_new_basis}
\end{eqnarray}
with 
\begin{eqnarray}
\hat{n}_a &=& \hat{z} \cos \theta_a + \hat{x} \sin
\theta_a = R_{\hat{y}} (\theta_a ) \hat{z} ,\nonumber\\
\hat{n}_f &=& \hat{y} \times \hat{n}_a
 \cos \delta_a  - \hat{y} \sin\delta_a = R_{\hat{z}'} (- \delta_a )
\hat{x}'  .
\end{eqnarray}
In place of Eq. (\ref{nt_euler}), we have
\begin{equation}
\pmb{\cal P}_t = \hat{n}_t = R_{\hat{y}''} (\theta_f ) R_{\hat{z}'} (-
\delta_a ) R_{\hat{y}} (\theta_a ) \hat{z} = R_{\hat{y}''} (\theta_f )
R_{\hat{y}} (\theta_a ) \hat{z} .
\label{nt_euler2}
\end{equation}  
By substituting the relations $\cot \theta_a = - \epsilon_a$ and $\cot
\theta_f = - \epsilon_{\rm BF,r}$ into Eq. (\ref{nt_euler2}), it is
checked that the same formula as Eq. (\ref{nt}) is obtained for
$\hat{n}_t$.  Note that the formula (\ref{nt}) for $\hat{n}_t$ is
independent of energy in contrast to $\hat{n}_a$ ($\hat{n}_f$) which
varies from $\hat{z}$ ($\hat{x}$) through $\hat{x}$ ($-\hat{z}$) to
$-\hat{z}$ ($-\hat{x}$) as energy varies from $-\infty$ to
$\infty$. This holds generally at least for the multichannel system in
the neighborhood of an isolated resonance and derives from the fact
that only one type of continua can interact with a discrete state (see
Eq. (\ref{QPa}) in Appendix \ref{App:deriv_na_nt} and
Ref. \cite{Lyuboshitz77} for more general systems).

So far, several coordinate systems are considered such as $xyz$,
$x'y'z'$, $x''y''z''$, and $x'''y'''z'''$ coordinate systems
pertaining to the eigenchannels of $S^0$, $S$ or $\boldsymbol{\sigma}
\cdot \pmb{\cal P}_a$, $\boldsymbol{\sigma} \cdot \pmb{\cal P}_f$, and
$\boldsymbol{\sigma} \cdot \pmb{\cal P}_t$, respectively.  These
coordinate systems are shown graphically in Fig. \ref{fig:euler}.
According to Eq. (\ref{theta_t}), the spherical polar coordinate of
$\hat{n}_t$ is given by (1,$\theta_t$, $-\Delta_{12}^0$) in the $xyz$
coordinate system.  Since $\hat{n}_a$ lies on the $zx$ plane, the
absolute magnitude $\Delta_{12}^0$ of the azimuth of $\hat{n}_t$ is
equal to the dihedral angle between two planes whose normals are given
by $\hat{z} \times \hat{n}_a$ and $\hat{z} \times \hat{n}_t$. Let us
now consider the coordinate of $\hat{n}_t$ in the $x'y'z'$
coordinates, where $\hat{z}'$ = $\hat{n}_a$. The angle which
$\hat{n}_t$ makes with $\hat{z}'$ is $\theta_f$ and the  azimuth of
$\hat{n}_t$ may be
obtained by considering the $z'x''$ (=$z''x''$) plane. Note that
$\hat{x}''$ is equal to $\hat{n}_f$ and $\hat{n}_t$ lies on the
$z'x''$ plane meaning that the azimuth of $\hat{n}_t$ is identical
with the dihedral angle which the $z'x''$ plane makes with the $z'x'$
plane.  Since $x''y''z''$ coordinate system is obtained from
$x'y'z'$ coordinate system by rotating about the $z'$ axis by
$-\delta_a$, the dihedral angle which the plane $z'x''$ makes with the
$z'x'$ plane is $\delta_a$. Therefore, the spherical polar coordinate
of $\hat{n}_t$ in the $x'y'z'$ coordinate system is (1, $\theta_f$,
$-\delta_a$). See Fig. \ref{fig:pol_vec} to understand the explanation
graphically.
Since the dihedral angle between two planes whose normals
are $\hat{n}_a \times \hat{x}'$ and $\hat{n}_a \times \hat{n}_t$ is
$\delta_a$, the dihedral angle between two planes whose normals are
given by $\hat{n}_a \times \hat{z}$ and $\hat{n}_a \times \hat{n}_t$
is $\pi - \delta_a$.  With this, we can construct a spherical triangle
$\Delta {\rm APQ}$ with vertices formed with the endpoints of
$\hat{z}$, $\hat{n}_a$, and $\hat{n}_t$, where the vertex angles
opposite to the edge angles $\theta_f$ and $\theta_t$ are
$\Delta_{12}^0$ and $\pi - \delta_a$, respectively, as shown in
Fig. \ref{fig:sph_tri}. The vertex angle opposite to the edge angle
$\theta_a$ can be shown to be $ \delta_r$ by making use of the
following relation
\begin{equation}
e^{-i \Delta_{12}^0 \boldsymbol{\sigma} \cdot \hat{z}} e^{-i \delta_r
\boldsymbol{\sigma} \cdot \hat{n}_t} = e^{-i \delta_a
\boldsymbol{\sigma} \cdot \hat{n}_a} ,
\label{na_nt}
\end{equation}
which is the spin model version of the relation ${\cal S}$ = ${\cal
S}^0 (\pi_b + 
e^{-2i \delta_r } \pi_a )$ (see Appendix \ref{App:deriv_na_nt} for the
derivation).  According to Appendix
\ref{App:mul_rot}, Eq. (\ref{na_nt}) can be expressed using the
rotation matrices in the Liouville space as 
\begin{equation}
R_{\hat{z}} (2 \Delta_{12}^0 ) R_{\hat{n}_t} (2 \delta_r ) =
R_{\hat{n}_a} (2 \delta_a )  
\end{equation}
and manifests that
the vertex angle
opposite to $\theta_a$ is $\delta_r$.

The dual spherical triangle of $\Delta {\rm APQ}$ may be constructed
by converting vertices of the original triangle to its edges and edges
of the original triangle to its vertices, according to the rule
described in Ref. \cite{JenningsBook}. According to the rule, the
vertex angles of the dual spherical triangle are obtained from the
corresponding edge angles of the original triangle by subtracting the
edge angles from $\pi$ like $\pi -\theta_a$, $\pi - \theta_f$ and $\pi
- \theta_t$ and the corresponding edge angles are also obtained
similarly like $\pi -\delta_r$, $\pi - \Delta_{12}^0$ and
$\delta_a$. The dual spherical triangle constructed in this way is
shown in Fig. \ref{fig:dual_st}.

Before considering dynamic aspects of the laws holding for the
spherical triangle, let us comment on Eq. (\ref{na_nt}), or the
equivalent ${\cal S}^0 (\pi_b + e^{-2i \delta_r} \pi_a )$. For this
purpose, let us define phase shift matrices $\boldsymbol{\Delta}^0$,
$\boldsymbol{\Delta}_r$, $\boldsymbol{\Delta}$ by ${\cal S}^0$ =
$e^{-2i \boldsymbol{\Delta}^0}$, $\pi_b + e^{-2i \delta_r} \pi_a$ =
$e^{-2i \boldsymbol{\Delta}_r}$, ${\cal S}$ = $e^{-2i
\boldsymbol{\Delta}}$. Phase shift matrices are easily obtained as
\begin{eqnarray}
\boldsymbol{\Delta}^0 &=& \frac{1}{2} (\delta_{\Sigma}^0 {\bf 1} +
\Delta_{12}^0 \boldsymbol{\sigma} \cdot \hat{z} ) ,
\label{Delta0} \\
\boldsymbol{\Delta}_r &=& \frac{1}{2} (\delta_r {\bf 1} + \delta_r
\boldsymbol{\sigma} \cdot \hat{n}_t ) ,
\label{Deltar} \\
\boldsymbol{\Delta}   &=& \frac{1}{2} ( \delta_{\Sigma}{\bf 1} + \delta_a
\boldsymbol{\sigma} \cdot \hat{n}_a ) .
\label{Delta}
\end{eqnarray}
For Eq. (\ref{Delta0}), if $\Delta_{12}^0$ = 0, two
eigenchannels have the identical background eigenphase shifts
$\delta_1^0$ = $\delta_2^0$ = $\delta_{\Sigma}^0 /2$. The
eigenchannels are isotropic with respect to the potential
that brings about the background phase shifts. When two
eigenchannels react differently or anisotropically with respect to the
potential, eigenchannels have different phase shifts $\delta_1^0$
$\ne$ $\delta_2^0$, or $\Delta_{12}^0$ $\ne$ 0. (The off-diagonal term
of $S^0$ gives the transition amplitude and is thus caused by the
channel-channel coupling. Since the off-diagonal term is implicitly
included in eigenchannels, the potential that eigenchannels feels
includes the channel-channel coupling effect.) The anisotropic term
of (\ref{Delta0}) contains the information on  this phase difference
and the eigenchannels. 

The phase shift matrix (\ref{Delta0}) is a vector whose coordinate is
($\delta_{\Sigma}^0,0,0,\Delta_{12}^0$) in the four-dimensional
Liouville space. Or, if we consider only the anisotropic term, it is a
vector in the three-dimensional Liouville space, whose magnitude is
$\frac{1}{2} \Delta_{12}^0$ and whose direction is $\hat{z}$.  Though
background and resonance scattering contributions appear as a single
product term in Eq. (\ref{na_nt}) in the $S$ matrix, two contributions
are not simply combined in the case of the phase shift matrix
$\boldsymbol{\Delta}$. For the isotropic parts of two contributions to
$\boldsymbol{\Delta}$, the combining rule is simple and they are
simply added up to give the isotropic part of the phase shift matrix
$\boldsymbol{\Delta}$ as $\frac{1}{2}\delta_{\Sigma}$ = $\frac{1}{2}
(\delta_{\Sigma}^0 + \delta_r)$. The combining rule of anisotropic
terms is not so simple. According to the Campbell-Baker-Hausdorff
formula\cite{Weiss62}, the anisotropic part of the phase shift matrix
$\boldsymbol{\Delta}$ is expressed as a very complicated infinite sum
of multiple commutators of the anisotropic parts of
$\boldsymbol{\Delta}^0$ and $\boldsymbol{\Delta}_r$ as
\begin{eqnarray}
2\boldsymbol{\Delta} - \delta_{\Sigma} {\bf 1} &=&
\Delta_{12}^0 \boldsymbol{\sigma} \cdot \hat{z} + 
\delta_r \boldsymbol{\sigma} \cdot \hat{n}_t -\frac{i}{2} 
[ \Delta_{12}^0 \boldsymbol{\sigma} \cdot \hat{z} ,
\delta_r \boldsymbol{\sigma} \cdot \hat{n}_t ]   \nonumber\\
&-& \frac{1}{12} 
\left( \left[  [ \Delta_{12}^0 \boldsymbol{\sigma} \cdot \hat{z} ,
\delta_r \boldsymbol{\sigma} \cdot \hat{n}_t ] ,
\delta_r \boldsymbol{\sigma} \cdot \hat{n}_t \right] -
\left[  [ \Delta_{12}^0 \boldsymbol{\sigma} \cdot \hat{z} ,
\delta_r \boldsymbol{\sigma} \cdot \hat{n}_t ] ,
 \Delta_{12}^0 \boldsymbol{\sigma} \cdot \hat{z} \right] \right)  +
\cdots 
\end{eqnarray}
But, the geometrical construction in the
Liouville space provides a simple combining rule. For that, at first
we ignore the magnitudes of vectors in the three-dimensional Liouville
space and consider the spherical
triangle made up of endpoints of the unit vectors corresponding to the
anisotropic terms. The magnitudes of vectors corresponding to
anisotropic terms, instead, are utilized as the edge angles of the
spherical triangle. This is the procedure we take when we interpret
Eq. (\ref{na_nt}) as giving the remaining edge angle $\delta_r$.
Trigonometric laws of the spherical triangle then provide a details of
the combining
rule which are the subject of the next section.

Let us comment one more thing on the avoided crossing interaction.  If
we use background eigenchannels as a basis, then the $S^0$ matrix is
already diagonal by definition in that basis. Then for processes
occurring along this eigenchannels, there is no channel-channel
coupling.  If a
discrete state is included into the system, background eigenchannels
are no longer decoupled and interact each other through the indirect
continuum-continuum, or channel-channel, coupling via the discrete
state. This indirect channel-channel coupling brings about the avoided
crossing interaction in the curves of eigenphase shifts of the $S$
matrix. Therefore, the avoided crossing interaction is not devoid of
the resonance contribution. What is devoid of in the avoided crossing
interaction is the isotropic resonant contribution. It includes the
anisotropic resonant contribution.

\section{Connection of the geometrical relation with dynamics}

Now let us describe the dynamical aspects of the geometrical laws
holding for the spherical triangle, such as the laws of sines, the
laws of cosines.  Cotangent laws including four
successive parts and laws including five successive parts (see
Ref. \cite{ChiangBook}) are derivable from the laws of sines and
cosines but deserve a treatment as separate laws. 

Let us consider the cotangent
law\cite{ChiangBook}
\begin{equation}
\sin \Delta_{12}^0 \cot \delta_a = -
\sin \theta_a \cot \theta_t + \cos \theta_a \cos \Delta_{12}^0.
\label{cot_delta_a}
\end{equation}
When $\cos \theta_a$ = $-\epsilon_a / \sqrt{\epsilon_a^2 +1}$ and
$\sin \theta_a$ = $1/\sqrt{\epsilon_a^2 +1}$ are inserted (for the
sign convection, see\cite{Convention2}), Eq. (\ref{cot_delta_a}) can be
put into Beutler-Fano's formula (\ref{delta_a_bf}) for $\cot \delta_a$
as follows
\begin{eqnarray}
\cot \delta_a &=& \frac{1}{\sin \Delta_{12}^0} \left(
\cos \theta_a \cos \Delta_{12}^0  - \sin \theta_a \cot
\theta_t \right) \nonumber \\ 
&=& - \frac{1}{\sin
\Delta_{12}^0} \left( \frac{\epsilon_a}{\sqrt{\epsilon_a^2 +1}} \cos
\Delta_{12}^0 + \frac{1}{\sqrt{\epsilon_a^2 +1}} \cot \theta_t \right)
\nonumber\\ 
&=& - \cot \Delta_{12}^0 \frac{\epsilon_a -
q_a}{\sqrt{\epsilon_a^2 +1}} ,
\end{eqnarray}
where $q_a$ is identified with $-\cot \theta_t / \cos \Delta_{12}^0$
and can be easily checked to be equal to the previous definition
(\ref{q_a}) with the use of Eq. (\ref{theta_t}).  

One of the cotangent law for its dual
spherical triangle is given by 
\begin{equation}
\cot \theta_f \sin \theta_t = - \cos (\pi - \delta_r ) \cos \theta_t +
\sin (\pi - \delta_r ) \cot \Delta_{12}^0 .
\label{cot_theta_f}
\end{equation}
With $\cos \delta_r$ = $-\epsilon_r /
\sqrt{\epsilon_r^2+1}$
and $\sin\delta_r$ = $1/\sqrt{\epsilon_r^2 +1}$,
Eq. (\ref{cot_theta_f}) 
can be
put into Beutler-Fano's formula (\ref{cotf}) for $\cot \theta_f$ as
follows
\begin{eqnarray}
\cot \theta_f &=& \frac{1}{\sin \theta_t} \left( \cos \delta_r \cos
\theta_t + \sin \delta_r \cot \Delta_{12}^0 \right) \nonumber\\
&=& - \frac{1}{\sin \theta_t} \left( \frac{\epsilon_r
}{\sqrt{\epsilon_r ^2 +1}} \cos \theta_t - \frac{1}{\sqrt{\epsilon_r
^2 +1}} \cot \Delta_{12}^0 \right) \nonumber\\
&=& -\cot \theta_t \frac{\epsilon_r -q_{\tau}} {\sqrt{\epsilon_r ^2
+1}} ,
\label{cotf_bf}
\end{eqnarray}
where $q_{\tau}$ is identified with $\cot \Delta_{12}^0 / \cos
\theta_t$, again equal to Eq. (\ref{qtau}). 

Using Eq. (\ref{cot_energy}) and the convention\cite{Convention2}, the
sine laws,
\begin{eqnarray}
\frac{\sin \Delta_{12}^0}{\sin \theta_f} &=& \frac{\sin \delta_r}
{\sin \theta_a} , \nonumber\\ 
\frac{\sin \delta_r}{\sin \theta_a} &=&
\frac{\sin \delta_a}{\sin \theta_t}
\label{sin_law}
\end{eqnarray}
are translated into 
\begin{eqnarray}
\epsilon_a ^2 +1 &=& \sin
^2 \Delta_{12}^0 (\epsilon_r^2 +1)(\epsilon_{{\rm BF,r}}^2 +1) ,
\label{ea2}
\\
\epsilon_r ^2 +1 &=& \sin ^2 \theta_t (\epsilon_a^2
+1)(\epsilon_{{\rm BF,a}}^2 +1),  
\label{er2}
\end{eqnarray}
respectively. Eq. (\ref{ea2}) was used to derive Eq. (\ref{tauf_bf})
in Ref. \cite{Lee98}. 

(Equating the $x,y,z$ components of $\hat{n}_t$ with those of
$\hat{n}_a \cos \theta_f + \hat{n}_f \sin \theta_f$ yields the law
involving 5 successive parts $\sin \theta_t \cos \Delta_{12}^0$ =
$\sin \theta_a \cos \theta_f$ + $\cos \theta_a \sin \theta_f
\cos\delta_a$, one of the laws of sines, $\sin \Delta_{12}^0 / \sin
\theta_f$ = $\sin \delta_a / \sin \theta_t$, and one of the laws of
cosines, $\cos \theta_t$ = $\cos \theta_a \cos \theta_f$ $-\sin
\theta_a \sin \theta_f \cos \delta_a$, in that order. Though such an
equality looks irrelevant to the laws of the spherical triangle at a
first glance, it actually has to do with the laws of the spherical
triangle since it is used to obtain the vertex angle $\pi -\delta_a$
of the endpoint of $\hat{n}_a$.)

The presence of the dual spherical triangle indicates that there is a
symmetry with respect to the exchange of vertices and edges. The
comparison of the spherical triangle in Fig. \ref{fig:sph_tri} with
its dual one in Fig. \ref{fig:dual_st} shows that the exchange of
$\delta_r$, $\theta_f$, and $\Delta_{12}^0$ with $\pi -\theta_a$,
$\delta_a$, and $\pi - \theta_t$ transforms the spherical triangle
into its dual and vice versa and thus any trigonometric laws will be
invariant under this exchange. Besides the geometrical laws, other
laws containing not only geometric parameters but also other types of
parameters should remain as valid expressions with respect to this
exchange. With this requirement, in order for $\cot \delta_r$ =
$-\epsilon_r$ to remain as a valid expression with respect to the
exchange, $\epsilon_r$ should be replaced by $-\epsilon_a$. The right
hand side of $q_{\tau}$ = $\cot \Delta_{12}^0 / \cos \theta_t$ becomes
$\cot \theta_t / \cos \Delta_{12}^0$ which is equal to $-q_a$. Thus
$q_{\tau}$ is replaced by $-q_a$ under the exchange. Similar procedure
shows that $\epsilon_{{\rm BF},a}$ is transformed into $\epsilon_{{\rm
BF},r}$ under the exchange. The variables with their conjugated ones
are summarized in Table \ref{table:conjug}. This symmetry under the
exchange yields many relations without derivation thus saves a lot of
efforts. It can also be used to check the validity of derived
equations. Let us take a few examples.  If it holds the relation that
\begin{equation}
\frac{\epsilon_r - q_{\tau}}{\sin \theta_t} = \frac{\epsilon_a +
\frac{1}{q_a}}{\sin \Delta_{12}^0} ,
\label{e_r_e_a}
\end{equation}
another valid relation is obtained as 
\begin{equation}
\frac{\epsilon_a - q_a}{\sin \Delta_{12}^0} = \frac{\epsilon_r +
\frac{1}{q_{\tau}}}{\sin \theta_t} ,
\end{equation}
by exchanging variables according to Table \ref{table:conjug}.
If it holds that
\begin{equation}
\frac{d \delta_a}{d\delta_r } = \cos \theta_f ,
\end{equation}
then the following
\begin{equation}
\frac{d\theta_f}{d\theta_a} = - \cos \delta_a 
\end{equation}
can be obtained by the same procedure. 

Geometrical realization reveals that complicated behaviors of
dynamical parameters like $\delta_a$ and $\theta_f$ as a function of
energy are nothing but the result of a simple geometrical traversal
along the great circle shown in Fig. \ref{fig:sphere}.  Before
examining the behaviors of dynamic parameters as functions of energy,
it is noted that the vectors $\hat{z}$ and $\hat{n}_t$ are fixed in
the real three-dimensional Liouville space while $\hat{n}_a$ changes
its direction as energy varies. The constancy of the $\hat{z}$ vector
derives from the usual assumption of energy insensitiveness of
background scattering. The constancy of the $\hat{n}_t$ vector derives
from that the time delay matrix has only a resonant contribution as
shown in Eq. (\ref{QPa}) and its eigenchannels consist of Fano's
energy independent $\psi_E^{(a)}$ and continua orthogonal to it.  As
the energy $\epsilon_r$ varies from $-\infty$ to $\infty$, $\theta_a$
undergoes a change from 0 to $\pi$ which corresponds to the
semicircular traversal of the point P from A to the opposite point
$-$A along the greatest circle while the points A and Q keeps fixed in
Fig. \ref{fig:sphere}. In this semicircular traversal, the angle $\pi
- \delta_a$ varies from $\pi - \Delta_{12}^0$ when P coincides with A
to $\Delta_{12}^0$ when P coincides with $-$A. The angle $\delta_a$,
accordingly, varies from $\Delta_{12}^0$ to $\pi -
\Delta_{12}^0$. (The angle $\theta_f$ varies similarly from $\theta_t$
to $\pi - \theta_t$.)  Traversal enjoys a special symmetry when
$\theta_t$ = $\pi/2$. Let the point A be taken as a polar point.  Then
the side $PQ$ becomes part of the equator at the middle of the
traversal where $\theta_a$ becomes $\pi/2$.  Since any meridian makes
the right angle with the equator, the angle $\delta_a$ which the chord
AP makes with the equator becomes a right angle, i.e., $\delta_a$ =
$\pi/2$.  Now let us consider the deviation of the point P from the
equator. Let $\delta_a$ = $\pi /2 + y$ at $\theta_a$ = $\pi /2
+x$. Then, by the symmetry of the spherical triangle, $\delta_a$ =
$\pi /2 - y$ at $\theta_a$ = $\pi /2 -x$. Napier's rule
\begin{equation}
\cot \delta_a = \cos \theta_a \cot \Delta_{12}^0
\label{cot_parallel}
\end{equation}
holding for the right spherical triangle satisfies such a
symmetry\cite{NapierRule}. Obviously, $q_a$ = 0 for
Eq. (\ref{cot_parallel}). When $\theta_t$ $\ne$ $\pi/2$, $\delta_a$ is
no longer $\pi /2$ when $\theta_a$ = $\pi /2$. The occurrence of the
mismatch in energies where values of $\theta_a$ and $\delta_a$ become
$\pi /2$ amounts to the addition of $\sin \theta_a$ term on the right
hand side of Eq. (\ref{cot_parallel}), which causes the value of $q_a$
to deviate from zero. The value of $q_a$ can be obtained as $-\cot
\theta_t /\cos \Delta_{12}^0$ by
substituting $- \cot \theta_t / \sin \Delta_{12}^0$, one of Napier's
rules holding when $\theta_a$ = $\pi /2$, for $\cot \delta_a$ into
Eq. (\ref{qaf_new_def}). It can be roughly stated that the asymmetry
of the Beutler-Fano formula for $\cot \delta_a$ derives from the
asymmetry of the geometry.

The concurrent change of $\theta_f$ with the increase of the arc
length $\theta_a$ as P traverses
can be obtained by differentiating the cosine law $\cos
\theta_f$ = $\cos \theta_a \cos \theta_t$ + $\sin \theta_a \sin
\theta_t \cos \Delta_{12}^0$ with respect to $\theta_a$ keeping
$\theta_t$ and $\Delta_{12}^0$ fixed, which becomes
\begin{equation}
- \sin \theta_f \frac{d \theta_f}{d \theta_a} = - \sin \theta_a \cos
\theta_t + \cos \theta_a \sin \theta_t \cos \Delta_{12}^0 .
\label{cos_deriv}
\end{equation}
The right hand side of Eq. (\ref{cos_deriv}) becomes $- \sin \theta_f
\cos ( \pi - \delta_a ) $ according to the law containing five
successive parts, which finally yields
\begin{equation}
\frac{d\theta_f}{d\theta_a} = - \cos \delta_a .
\end{equation}
Similar derivatives are obtained as
\begin{eqnarray}
\frac{d \delta_a}{d\delta_r}  &=& \cos \theta_f ,\nonumber\\
\frac{d \delta_a}{d \theta_a} &=&  \cot \theta_f \sin \delta_a ,
\nonumber\\
\frac{d \cot \theta_a}{d \cot \delta_r}  &=&
   \frac{\sin \Delta_{12}^0}{\sin \theta_t} .
\label{angle_deriv}
\end{eqnarray}

In the spherical triangle $\Delta ABC$, Gauss-Bonnet theorem becomes
\begin{equation}
\angle A + \angle B + \angle C = \pi + \frac{area(\Delta ABC)}{R^2} ,
\end{equation}
which states that the sum of interior angles of a spherical triangle
exceeds $\pi$ by the solid angle $\Omega$ defined by $area(\Delta
ABC)/R^2$. In the present case, the sum of interior (vertex) angles is
$\pi + \delta_r + \Delta_{12}^0 - \delta_a $. Hence, the solid angle
$\Omega$ is
\begin{equation}
\Omega = \delta_r + \Delta_{12}^0 - \delta_a = \delta_r +
\delta_{\Sigma}^0 -2 \delta_2^0 - \delta_a = 2 (\delta_{-} -
\delta_2^0 ) .
\label{Omega}
\end{equation}
The solid angle of the spherical triangle $\Delta APQ$ is easily
calculated as $2 \Delta_{12}^0$ when the point P coincides with the
antipode of A.  Then the solid angle of the spherical triangle varies
from zero to $2 \Delta_{12}^0$ as the point P varies from the point A
to the point $-$A and, accordingly, 
$\delta_{-}$ varies from $\delta_2^0$
to $\delta_1^0$ as energy varies from $-\infty$ to $\infty$, which is
consistent with the result of Ref. \cite{Lee98}.

So far, a geometric realization of the $S$ matrix and $Q$ matrix has
been considered. Let us now go back to the original questions we had
in the beginning of Sec. \ref{Sec:Prep} and see whether we can explain
them.

The first of the questions was that the energy variations of the
eigenvectors of the $S$ matrix are independent of $q_a$ while those of
its eigenphase shifts, or more specifically $\delta_a$, depend on
it. Let us start with that the eigenvectors or eigenchannels of the
$S$ matrix are obtained by the frame transformation of the background
eigenchannels. Since the background eigenchannels are fixed, the
energy variations of eigenchannels of the $S$ matrix completely come
from the frame transformation which in this case is parametrized with
$\theta_a$. In Fig. \ref{fig:euler}, $\theta_a$ is the edge angle
opposite to the vertex angle $\delta_r$. If $\epsilon_r$ varies,
$\delta_r$ varies according to $\cot \delta_r$ = $-\epsilon_r$. Since
the edge angle $\theta_a$ is the opposite to $\delta_r$, $\theta_a$
may be expected to vary linearly with $\delta_r$. Such an expectation
does not come out right. Instead, $\cot \theta_a$ varies linearly with
$\cot \delta_r$ according to one of the relations in
(\ref{angle_deriv}), $d \cot \theta_a / d \cot \delta_r$ = $\sin
\Delta_{12}^0 / \sin \theta_t$, which is fixed in energy. The relation
tells us that $\cot \theta_a$ has a linear relation with $\epsilon_r$
but $\cot \theta_a$ may not be zero when $\epsilon_r$ = 0, in
general. But, we can always introduce a new energy scale, let us call
it $\epsilon_a$, where $\cot \theta_a$ is zero at $\epsilon_a$ = 0 and
the proportionality constant can be set so that $\cot \theta_a = -
\epsilon_a$. The argument proves that $\cot \theta_a$ does not need no
further parameter like $q_a$.  The reason why the energy variation of
$\delta_a$ needs the line profile index is already considered around
Eq. (\ref{cot_parallel}) and need not be repeated here.

If the second question which asks why the energy behaviors of
$\delta_a$, $\tau_a$ and $\tau_f$ follow the Beutler-Fano formulas is
changed like ``is it possible to show geometrically that their
behaviors follow the Beutler-Fano formula?'', the answer is yes and
their behavior is the result of the cotangent laws holding for the
spherical triangle, as we have done in this section.
 
Let us answer the third question why $\tau_a$ takes the Beutler-Fano
formula in the energy scale of $\epsilon_r$ instead of $\epsilon_a$
though $\tau_a$ is obtained as the derivative of $\epsilon_a$. 
Note that the question on
$\tau_a$ can be paraphrased to that on $\cot \theta_f$ since $\tau_a$
= $\tau_r (\epsilon_r ) (1+\cot \theta^2_f )^{-1/2}$.
There are two such cotangent laws for $\cot \theta_f$ as
follows
\begin{eqnarray}
\cot \theta_f &=& \frac{1}{\sin \theta_t} \left(\cos \delta_r \cos
\theta_t + \sin \delta_r \cot \Delta_{12}^0 \right) , 
\label{cotf_delta_r}
\\
\cot \theta_f &=& \frac{1}{\sin \theta_a} \left(- \cos \delta_a \cos
\theta_a + \sin \delta_a \cot \Delta_{12}^0 \right) .
\label{cotf_delta_a}
\end{eqnarray}
Eq. (\ref{cotf_delta_r}) expresses $\cot \theta_f$ in terms of
$\epsilon_r$ while Eq. (\ref{cotf_delta_a}) expresses it in terms of
$\epsilon_a$.  In Eq. (\ref{cotf_delta_r}), $\delta_r$ is the only
parameters which is a function of energy while, in
Eq. (\ref{cotf_delta_a}), not only $\delta_a$ but also $\theta_a$ are
parameters which are functions of energy. Eq. (\ref{cotf_delta_r})
gives the Beutler-Fano formula as a function of $\epsilon_r$ as we
already saw in Eq. (\ref{cotf_bf}). Eq. (\ref{cotf_delta_a}) might
also give a Beutler-Fano formula as a function of $\epsilon_a$ if
$\theta_a$ were a constant of energy but it fails to do so as
$\theta_a$ is a function of energy, too. On the other hand, if
$\theta_t$ were a function of energy, Eq. (\ref{cotf_delta_r}) could
not give the Beutler-Fano formula, too. This argument reveals that
cotangent laws of a spherical triangle alone are not sufficient to
guarantee the presence of Beutler-Fano formulas.

Let us consider the answer to the fourth question which asks the
reason why $| {\bf P}_t |^2$ = $|{\bf P}_a |^2 + |{\bf P}_f |^2$ = 1
is satisfied.  Eq. (\ref{QPa}) shows that non-zero resonant behavior
of time delay occurs
only when the system is in the $\psi_E^{(a)}$ state, which derives
from that the $\psi_E^{(a)}$ state is the only type of continuum which
can interact with the discrete state. On the other
hand, the unit magnitude of the polarization ${\bf P}_t$ means that
only one continuum shows a resonant behavior while others do not.
Thus Eq. (\ref{QPa}) proves that $|{\bf P}_t|$ = 1.

The answer to the fifth question is provided by the identification of
$r^2$ with $\tan \theta^2 _t$ and need not be considered further.

\section{Application to the triatomic van der Waals
predissociation \label{sec:apps}}

Ref. \cite{Lee98} and the present paper have developed the theory for
the behaviors of eigenphase shifts and time delays.  Let us now
consider the application of the theory to the vibrational
predissociation of triatomic van der Waals molecules. The theory can
be applied in two ways. When the $S$ matrix is known as a function of
energy either experimentally or by a theoretical calculation,
eigenphase shifts can be calculated directly by its
diagonalization. Similarly, eigentime delays can be calculated from
the $S$ matrix. For these data, formulas for eigenphase shifts and
time delays derived from the theory can be used as models with
parameters in the formulas viewed as adjustable ones. Extracting best
values of the parameters may be tried by fitting the data of
eigenphase shifts and eigentime delays obtained by the diagonalization
of the $S$ matrix to the theoretical models.  On the other hand, by
using the formulas for the parameters themselves derived from the
theory, parameters can be directly calculated without doing data
fitting.  Parameters obtained in two different ways, namely, by
data-fitting and by using the theoretical formulas are not identical
as the theory developed so far relies on the assumption that
background eigenphase shifts and partial decay widths are constants of
energy, which is usually a good approximation but does not hold
exactly in the actual system.

The data-fitting will be done only for the eigentime delay sum
(\ref{eigentime_delay_sum}) and
partial delay times $2\hbar d\delta_{+}/dE$ and $2\hbar d\delta_{-}
/dE$. Eigenphase shifts will not be used for the data-fitting since
they need the information on $E_a$ and $\Gamma_a$, which is not
available before the data-fitting. The data-fitting of partial delay
times to the theoretical formulas
\begin{eqnarray}
2\hbar \frac{d\delta_{\pm}}{dE} &=& \hbar \left( \frac{d\delta_r}{dE}
\pm \frac{d\delta_a}{dE} \right) = \frac{1}{2} ( \tau_r \pm \tau_a ) =
\frac{1}{2} \tau_r (1 \pm \cos \theta_f ) \nonumber\\
&=& \frac{2\hbar}{\Gamma} \frac{1}{1+ \epsilon_r^2} 
\left[ 1 \mp \frac{\epsilon_r - q_{\tau}}{\sqrt{(\epsilon_r -
q_{\tau})^2 + \tan ^2 \theta_t (1+\epsilon_r^2 )}} \right]
\label{partial_delay_times}
\end{eqnarray}
can be easily done on the other hand since the information on $E_0$
and $\Gamma$ which are necessary to convert $E$ to $\epsilon_r$ needed
for Eq. (\ref{partial_delay_times}) can be easily
obtained from the data-fitting of the eigentime delay sum
(\ref{eigentime_delay_sum}). (Eigenphase sum can also be used to get
$E_0$ and $\Gamma$.)
       
Graphs of partial delay times  are shown
in Fig.  \ref{fig:dpm_df} for several values of line profile indices.  
Some general characteristics  of the graphs are noticed.
\begin{enumerate}
\item  Graphs of partial delay times $2 \hbar d\delta_{+} / dE$ and
$2\hbar d
\delta_{-} /dE$ meet at   $\epsilon_r$ = $q_{\tau}$. 
\item As  $ | q_{\tau} |$ $\rightarrow$ $\cot
\Delta_{12}^{\circ}$, 
\begin{equation}
\cos \theta_f (\epsilon_r ) \rightarrow
\left\{ \begin{array}{rl}
 1& {\rm when}~\epsilon_r \le q_{\tau} \\
-1& {\rm when}~\epsilon_r > q_{\tau} ,
\end{array} \right.
\label{cosf_qt_limit}
\end{equation}
and the partial delay times becomes 
\begin{eqnarray}
2\hbar \frac{d\delta_{+}}{dE} \rightarrow
\left\{ \begin{array}{ll} 
\tau_r (\epsilon_r ) &{\rm when}~\epsilon_r \le q_{\tau} \\
0 &{\rm when}~\epsilon_r > q_{\tau} , 
\end{array} \right.
\nonumber \\
2\hbar \frac{d\delta_{-}}{dE} \rightarrow
\left\{ \begin{array}{ll} 
0 &{\rm when}~\epsilon_r \le q_{\tau} \\
\tau_r (\epsilon_r ) &{\rm when}~\epsilon_r > q_{\tau}  .
\end{array} \right.  
\label{td_theta_t_zero}
\end{eqnarray}
This case corresponds to $\theta_t$ $\rightarrow$ 0, or $\theta_f$
$\rightarrow$ $\theta_a$, as  can be easily seen from the inspection of
Fig. \ref{fig:sphere}. But with this geometrical consideration alone,
it is hard to find the limit of $\cos \theta_f$  in
Eq. (\ref{cosf_qt_limit}). The behavior of the limit of $\cos
\theta_f$ can only be obtained when $\Gamma_a \rightarrow 0$ is taken
into account at $\theta_t \rightarrow 0$, i.e., when the strength of
the channel coupling is taken into account, which is hidden in
the geometrical consideration because of the use of $\epsilon_a$. When
$\Gamma_a \rightarrow 0$,
\begin{equation}
\epsilon_a \rightarrow
\left\{ \begin{array}{rl}
 - \infty & {\rm when}~E < E_a \\
\infty & {\rm when}~E > E_a . 
\end{array} \right.
\end{equation}
Only two values of $\epsilon_a$ are possible in the limit of
$\Gamma_a$ $\rightarrow$ 0 or $| q_{\tau} |$ $\rightarrow$ $\cot
\Delta_{12}^0$.  $\epsilon_a \rightarrow \mp \infty$ correspond to
$\theta_a$ $\rightarrow$ 0 and $\pi$, or $\cos \theta_a $
$\rightarrow$ 1 and $-$1, respectively. Since $\cos \theta_f$
$\rightarrow$ $\cos \theta_a$, $\cos \theta_f$ satisfies the limit
(\ref{cosf_qt_limit}) as $|
q_{\tau}|$ $\rightarrow$ $\cot \Delta_{12}^0$.  The energies at which
$\cos \theta_f$ and $\epsilon_a$ undergo abrupt changes look different
but are equivalent since $E = E_a$ or $\epsilon_a =0$ corresponds to
$\epsilon_r$ = $q_{\tau}$ owing to the relation (\ref{e_r_e_a}) and
$q_a \rightarrow \infty$.

\item As $| q_{\tau}| \rightarrow \infty$, the graph of $\cos
\theta_f (\epsilon_r )$ becomes symmetric with respect to origin and
is given  by
\begin{equation}
\cos \theta_f (\epsilon_r ) \rightarrow
 \frac{\cos \Delta_{12}^{\circ}}{\sqrt{\epsilon_r^2 
    \sin ^2 \Delta_{12}^{\circ}+1}} ~~~~~~~~{\rm when}~ |q_{\tau}|
\rightarrow \infty .
\label{cosf_q_infty}
\end{equation}
The derivation of Eq. (\ref{cosf_q_infty}) from
Eq. (\ref{partial_delay_times}) is not so easy.  $|q_{\tau}|$
$\rightarrow$ $\infty$ arises in two cases, $\cot \Delta_{12}^0$
$\rightarrow$ $\infty$ or $\cos \theta_t$ = 0 ($\theta_t$ =
$\pi/2$). The geometric consideration is a great help when $\theta_t$
= $\pi /2$. In this case, Napier's rule gives $\cot \theta_f$ =
$\sin\delta_r \cot \Delta_{12}^0$\cite{NapierRule}. From this formula
of $\cot \theta_f$, $\cos ^2 \theta_f$ = $\cos ^2 \Delta_{12}^0 /
(\epsilon_r^2 \sin ^2 \Delta_{12}^0 +1)$ is obtained by simple
trigonometric manipulations. The square roots of both sides of it
yields Eq. (\ref{cosf_q_infty}) except for the sign.  In order to fix
the sign, let us consider the case of $\epsilon_r$ = 0 which
corresponds to $\delta_r$ = $\pi/2$. Since $\theta_t$ = $\delta_r$ =
$\pi/2$ means that the chord PQ in Fig. \ref{fig:sphere} is part of
the equator, we have $\theta_a$ = $\delta_a$ = $\pi /2$. For this
particular spherical triangle, it can be easily proved that $\theta_f$
= $\Delta_{12}^0$. This fixes the sign. The remaining case of $\cot
\Delta_{12}^0$ $\rightarrow$ $\infty$ corresponds to $\Gamma_a$
$\rightarrow$ 0 and can not be easily handled by geometric argument as
mentioned above. Eq. (74) of Ref. \cite{Lee98} allows us to handle
this case and yields $\cos \theta_f$ = $\cos \Delta_{12}^0$  which is
identical with 
Eq. (\ref{cosf_q_infty}) in this case.

The partial delay times at $\epsilon_r$ = 0 are
\begin{eqnarray}
2\hbar \frac{d\delta_{+}}{dE}( \epsilon_r =0 ) &\rightarrow&
  \frac{4\hbar}{\Gamma} \cos ^2\left(\frac{\Delta_{12}^{\circ}}{2}
  \right) , \nonumber \\ 
2\hbar \frac{d\delta_{-}}{dE}( \epsilon_r =0 )
  &\rightarrow& \frac{4\hbar}{\Gamma} \sin
  ^2\left(\frac{\Delta_{12}^{\circ}}{2} \right) ,
\end{eqnarray}
which is easily obtained by substituting $\Delta_{12}^0$ for
$\theta_f$ into $2 \hbar d\delta_{\pm}/dE$ = $\tau_r (1\pm \cos
\theta_f )$.
\end{enumerate}

Before doing the data-fitting, let us briefly describe the system
used for the calculation and the methods of calculation.  A triatomic
van der Waals molecule considered here is of the type of rare
gas$\cdots$homonuclear halogen-like diatomic
molecules\cite{Delgado-Barrio95}. Let us consider the system where the
van der Waals molecule in its ground state is excited by the light
whose energy amounts to the excitation of the diatomic vibronic motion
from the $v$ = 0 to 1 state. This energy is sufficient to break the
van der Waals bond and produces a predissociation spectrum as the
energy of  light is scanned over a certain range of frequency.

The following interaction potential between A and B$_2$ in
A$\cdots$B$_2$ triatomic system
\begin{equation}
V(R,r,\gamma)= \left\{ \begin{array}{ll} V_{\rm M} (R,r,\gamma) &
{\rm when}~ R\le R^*  \\ 
V_{\rm vdW}(r,\gamma)+(V_{\rm M} -V_{\rm
vdW}) e^{-\rho \left( \frac{R-R^*}{R^*} \right) ^2}  & {\rm when}~R
\ge R^* ,
\end{array} \right.
\label{V_Jacobi}
\end{equation} 
is the one employed by Halberstadt et. al. to fit the predissociation
data for Ne$\cdots$Cl$_2$ system and is used here\cite{Halberstadt87}.
In Eq. (\ref{V_Jacobi}), $R,r,\gamma$ are the Jacobi coordinates that
denote the distance between A and the center of mass of B$_2$, the
bond distance of B$_2$, and the angle between ${\bf R}$ and ${\bf r}$,
respectively\cite{Beswick81}; $V_{\rm M} (R,r,\gamma) $ and $V_{\rm
vdW}$ are given as
\begin{equation} V_{\rm M} (R,r,\gamma)=D_{\rm AB} \sum_{i=1}^2 
\left\{ \left[
e^{-\alpha_{\rm AB}(R_{{\rm AB}_i}-R_{\rm AB}^{(o)})}-1 
\right] ^2-1 \right\} 
^2\end{equation} 
\begin{equation}
 + D_{\rm CM} \left\{ \left[ e^{-\alpha_{\rm CM}
(R-R_{\rm CM}^{(o)})} 
-1 \right] ^2 -1 \right\} ^2 , \end{equation} 
\begin{equation} V_{\rm vdW} (R,\gamma)= - { C_6
(\gamma ) \over R^6} - {C_8 (\gamma ) \over R^8 }  ,  
\label{Vvdw}
\end{equation} 
where $R_{{\rm AB}_i}$ is the distance between A and $i^{\rm th}$ B
atom; other parameters are adjusted parameters
to yield the best fit to the experimental values. The values
of the parameters used in this paper are given in Table \ref{potential}.
Two Legendre terms are retained for $C_6 (\gamma)$ and $C_8 (\gamma)$
in Eq. (\ref{Vvdw}),
e.g.,
\begin{equation} 
C_6 (\gamma ) = C_{60} +C_{62} P_2
(\cos\gamma ) . 
\end{equation} 
$R^*$ in Eq. (\ref{V_Jacobi}) is chosen as the inflection point of the 
atom-atom Morse potentials and given by $R^*=R_{\rm CM}^{(o)} +{\rm 
ln}2/\alpha_{\rm CM}$.  

The Hamiltonian for the triatomic van der Waals molecules
A$\cdots$B$_2$ in the Jacobi coordinates is given in atomic units
by\cite{Halberstadt87}
\begin{equation}
H= - {1 \over 2m} {\partial ^2 \over \partial R^2} + {{\bf j} ^2 \over
2 \mu r^2} + { {\bf l} ^2 \over 2m R^2} + V(R,r,\gamma ) + H_{\rm B_2}
(r), \label{H}
\end{equation} 
where 
\begin{equation}
H_{\rm B_2}(r) = -{1 \over 2\mu r^2 }{\partial ^2 \over \partial r^2} 
+ V_{\rm B_2} (r) ,
\end{equation} 
denotes the vibrational Hamiltonian of B$_2$; $m$ is the reduced mass
of B$_2$; $\mu$ denotes the reduced mass of A and the center of mass
of B$_2$; ${\bf j}$ is the angular momentum operator of B$_2$; ${\bf
l}$ is the orbital angular momentum operator of the relative motion of
A and the center of mass of B$_2$.  The values of diatomic molecular
parameters of B$_2$ used in this paper are given in Table
\ref{diatom}.  The calculation is limited to zero of the total angular
momentum operator ${\bf J} ={\bf j} +{\bf l}$, as usually done in this
field without affecting the predissociation dynamics much.  Such a
limitation simplifies the Hamiltonian as ${\bf l}$ can be replaced by
${\bf j}$.

Let $\Psi^{-(i)} (R,r,\gamma )$ be the eigenfunctions of $H$ of
Eq. (\ref{H}) and let it correspond to the state vibronically excited
by light which will be predissociating into an atom and a diatomic
fragment. It is indexed by the vib-rotational quantum numbers
($v,j$) of its diatomic photo-fragment and abbreviated
to $i$, i.e., $i$ = ($v,j$).  When the wavefunctions
$\Psi^{-(i)} (R,r,\gamma )$ to the dissociation channel
$i$ = ($v,j$) are expanded in terms of $n$ base functions
$\Phi_{i'} (r,\gamma )=(r|v')Y_{j' o} (\gamma,0)$ ($i'$ = 1,2,...,$n$)
as
\begin{equation}
\Psi^{-(i)} (R,r,\gamma )=\sum_{i'} \Phi _{i'} (r, \gamma ) \chi_{i'i}
(R),
\label{psi_vdw_asym}
\end{equation} 
the close-coupling equations for $\chi_{i'i} (R)$ are obtained as
\begin{equation} 
\left[ -{1 \over 2m} {d^2 \over dR^2}- k_{i'}^2 +{{\bf j} ^2 \over 2m
R^2 } \right] \chi _{i'i} (R) +\sum_{i''} V_{i'i''} (R)\chi_{i''i}
(R)=0, 
\label{cc}
\end{equation} 
with 
\begin{equation} 
k_{i'}^2 =
2m[E-Bj'(j'+1)-(v'+\frac{1}{2} )\omega ], 
\end{equation} 
and 
\begin{equation} 
V_{i''i'} (R)=\int d\gamma \sin\gamma \int dr \Phi_{i''}(r,\gamma
)V(R,r,\gamma )\Phi _{i'}^* (r,\gamma ) .
\end{equation} 

The close-coupling equation (\ref{cc}) is solved by the De Vogelaere
algorithm\cite{Lester71} and wavefunctions (\ref{psi_vdw_asym}) that
satisfy the incoming wave boundary condition are, consequently,
obtained. The $S$ matrix obtained in this process, which is identical
with (\ref{S_res}), is diagonalized and eigenphase shifts
(\ref{eigenphase_shifts}) are obtained. Two closed channels
corresponding to ($v=1,j=0$) and ($v=1,j=2$) and two open channels
corresponding to ($v=0,j=0$) and ($v=0,j=2$) are included to mimic
the system of one discrete state and two continua to which the theory
developed in this work applies.  This calculation
yields the data of eigenphase shifts as functions of energy. Let us
call this method of calculation as the close-coupling method.

Note that the theory developed in Ref. \cite{Lee98} and in the present
paper relies on the presence of a discrete state embedded in
continua. Among various theories devised to describe the resonance
with explicit consideration of a discrete state, Fano's configuration
interaction theory is chosen in this work\cite{Lee98}.  In the normal
use as described in the above paragraph, the close-coupling method can
not be connected with the configuration interaction theory since no
discrete state is assumed in the close-coupling method. But with a
little modification in its use, it can be used to calculate dynamic
parameters of the configuration interaction theory.  A discrete state
with its resonance energy $E_0$ used in the configuration interaction
theory can be obtained by solving the close-coupling equations
(modified to incorporate the shooting method\cite{NumericalBook}) with
inclusion of closed channels alone. Wavefunctions obtained by solving
the close-coupling equations with inclusion of open channels alone
obviously diagonalize the Hamiltonian in the subspace spanned by open
channels alone and are the continuum wavefunctions $\psi_E^{-(l)}$
considered in the configuration interaction theory.  The background
scattering matrix $S^0$ are obtained as a byproduct when the continuum
wavefunctions $\psi_E^{-(l)}$ are forced to satisfy the incoming wave
boundary conditions (or outgoing wave boundary condition if a
scattering system is considered instead of the photo-dissociation). By
diagonalizing $S^0$, background eigenphase shifts $\delta_1^0$ and
$\delta_2^0$ and the frame transformation matrix $U^0$ from the
asymptotic wavefunctions $\psi_E^{-(l)}$ to background eigen channel
wavefunctions $\psi_E^{(k)}$ are obtained. With $U^0$, background
eigenchannel wavefunctions can be obtained from the asymptotic ones as
\begin{equation}
\psi_E^{(k)} (R,r,\gamma ) = -i e^{i \delta_k^0} \sum_l
\tilde{U}_{kl}^0 \psi _E^{-(l)}(R,r,\gamma ) ,
\end{equation}
and can be used to calculate partial decay width $\Gamma_k$ as 
\begin{equation}
\Gamma_k = 2 \pi \left| \left( \psi_E^{(k)} | H | \phi \right) \right|^2
         = 2 \pi \left|\left( \psi_E^{(k)} | V(R,r,\gamma ) | \phi
         \right)\right| ^2 ,
\label{Gamma_k}
\end{equation}
where the last equality holds for $\delta v$ = $\pm 1$ vibronic
predissociation since $V(R,r,\gamma )$ is the only term containing odd
powers of $r$ in $H$. Since $E_0$, $\delta_1^0$, $\delta_2^0$,
$\Gamma_1$, and $\Gamma_2$ are obtained, the dynamic parameters $E_a$,
$\Gamma_a$, $q_{a}$, $q_{\tau}$, and $\cot \theta_t$ are calculated
using $E_a$ = $E_0$ + $\Delta \Gamma \cot \Delta_{12}^0 /2$,
$\Gamma_a$ = $2\sqrt{\Gamma_1 \Gamma_2} / \sin \Delta_{12}^0$,
Eqs. (\ref{q_a}), (\ref{qtau}), and (\ref{theta_t}).

Though the configuration interaction theory directly calculates the
dynamic parameters $E_0$, $\Gamma$, $E_a$, $\Gamma_a$, $q_a$,
$q_{\tau}$, and $\cot \theta_t$, the close-coupling method can not
calculate them directly.  Dynamical parameters directly obtainable
from the close-coupling method are the $S$ matrix, its eigenphase
shifts $ \delta_{+}$ and $ \delta_{-}$, and partial delay times
$2\hbar d\delta_{+} /dE$ and $2\hbar d\delta_{-} /dE$ as functions of
energy.  If the assumptions used in configuration interaction theory
are exact, eigenphase shifts and partial delay times should satisfy
Eq. (\ref{eigenphase_shifts}) and Eq. (\ref{partial_delay_times}),
respectively. The assumptions that $\delta_1^0$, $\delta_2^0$,
$\Gamma_1$, and $\Gamma_2$ are constant of energy on which the
configuration interaction theory relies are expected not to cause much
trouble in the actual situation as far as the energy range is not wide
enough. If eigenphase shifts and partial delay times follow
Eq. (\ref{eigenphase_shifts}) and Eq. (\ref{partial_delay_times}),
values of dynamic parameters $E_a$, $\Gamma_a$, $q_a$, $q_{\tau}$, and
$\cot \theta_t$ can be obtained by varying them so that eigenphase
shifts and partial delay times calculated by the close-coupling method
fit the formulas best.  Fitting is done with the Levenberg-Marquardt
method recommended for the nonlinear models in
Ref. \cite{NumericalBook}. For the reason mentioned above, only
partial delay times will be fitted.

Data fittings are done in two steps. At first, $E_{\circ}$ and
$\Gamma$ are obtained by fitting the eigentime delay sum. Then
$q_{\tau}$ and $\cot \theta_t$ are obtained by fitting partial delay
times $2\hbar d \delta_{+} /dE$ (=$\tau_r + \tau_a$) and $2\hbar
d\delta_{-} / dE$ (=$\tau_r - \tau_a$) with
(\ref{partial_delay_times}).  Either $2\hbar d\delta_{+} /dE$ or
$2\hbar d\delta_{-} /dE$ can be used to obtain the values of
$q_{\tau}$ and $\cot \theta_t$. Values of $q_{\tau}$ and $\cot
\theta_t$ obtained from either of them should be identical if the
assumption of the configuration interaction method, namely that
$\Gamma_1$, $\Gamma_2$, $\delta_1^0$, and $\delta_2^0$ are constant of
energy, is exact.  The differences between $q_{\tau}$'s and
$\cot\theta_t$'s for $2\hbar d\delta_{+} /dE$ and $2\hbar d\delta_{-}
/dE$ may serve as a criterion of the exactness of the configuration
interaction theory.

Numerical study shows that fitting the eigentime delay sum calculated
by the close-coupling method to the formula
(\ref{eigentime_delay_sum}) is done reliably. Compared with that,
fitting the partial delay times to the formula
(\ref{partial_delay_times}) can not be easily done. Eigentime delays
calculated by the close-coupling method show abnormal behaviors like
negative values of eigentime delays at the energy region not far from
a resonance while theory contends that eigentime delays are positive
in the neighborhood of a resonance, meaning that assumptions made for
the formula (\ref{partial_delay_times}) are prone to break down.  It
may be argued that parameters obtained by data-fitting of eigentime
delays calculated by the close-coupling method will deviate more from
those obtained by the configuration interaction method as the range of
energy taken for fitting is wider.  But we can not always narrow the
range of energy for this reason. Notice that $q_{\tau}$ is the energy
where $\tau_{+}$ meets $\tau_{-}$.  As seen in
Eq. (\ref{td_theta_t_zero}), if $q_{\tau}$ is large, two curves are
almost identical with $\tau_{r}$ except for the neighborhood of
$q_{\tau}$. Therefore for the good fitting, the energy range taken for
fitting should include $q_{\tau}$.  It means that if $q_{\tau}$ is
large, fitting is likely to be bad.  This assertion is checked by
doing data-fittings
for three different ranges of energy, namely, $[ E_{\circ}
-\Gamma, E_{\circ} +\Gamma ]$, $ [E_{\circ} -2\Gamma, E_{\circ}
+2\Gamma ]$, and $[ E_{\circ} -3\Gamma, E_{\circ} +3\Gamma ]$.

Table \ref{tab:fit3044} shows the results calculated with the
parameters given in Table \ref{potential} and \ref{diatom}. Table
\ref{tab:fit5044} is obtained with the same parameters but with $r_e$
= 5.044 a.u. The former table corresponds to the case of large
$q_{\tau}$ while the latter table to the case of small $q_{\tau}$.
The tables also show values of a fudge factor $\lambda$ which can be
used as a criteria for the goodness-of-fit. The fitting is done at
first by steepest decent method with the initial value, say 0.001, of
$\lambda$. New value of $\lambda$ is suggested for the next
iteration. If the suggested value of $\lambda$ becomes sufficiently
small, inverse-Hessian method is used for fitting. At the final call,
$\lambda$ is set to zero. This method assumes that values of $\lambda$
should approach zero if everything goes O.K. This is true with the
data-fitting of the eigentime delay sum. For the partial delay times,
values of $\lambda$ do not go to zero. Though values of $\lambda$ do
not go to zero, comparison of curves obtained by two method showed
that the fitting is good enough to be acceptable if values of
$\lambda$ are not too large. In Table \ref{tab:fit3044}, the values of
$q_{\tau}$ lies beyond the first interval [$E_{\circ} - \Gamma$,
$E_{\circ} +\Gamma$] and lies in the second interval. According to the
reasons mentioned above, it is hard to achieve reliable data-fitting
in this case. The closest result to the theoretical values is obtained
for the first interval, the narrowest one, where the fudge factor is
worst indicating wider interval should be used. The calculation shows
that wider interval yields worse result indicating that the assumptions
may no longer be true. On the other hand, in Table V, the value of
$q_{\tau}$ is small and the data-fitting may be done rather reliably
and is confirmed by the calculation. This situation contrasts greatly
to the fitting of partial photo-dissociation cross-sections to the
Fano-Beutler-like line profile formulas\cite{Lee95}
\begin{equation}
\sigma_j = \sigma_j ^{\circ} 
\frac{| \epsilon + q_j |^2}{1+\epsilon ^2} = 
\sigma_j ^{\circ} \frac{[ \epsilon + \Re (q_j ) ]^2}{1+\epsilon ^2} +
\sigma_j ^{\circ} \frac{[ \Im (q_j ) ]^2}{1+\epsilon ^2}  ,
\end{equation}
for the same predissociating system of van der Waals
molecules, where the fitting is excellent as shown in Table
\ref{tab:partial_photodissociation_cross_section}. 

\section{Summary and Discussion}

In the previous work\cite{Lee98}, eigenphase shifts for the $S$ matrix
and Smith's lifetime matrix $Q$ near a resonance were expressed as
functionals of the Beutler-Fano formulas using appropriate
dimensionless energy units and line profile indices. Parameters
responsible for the avoided crossing of eigenphase shifts and
eigentime delays and the change in frame transformation in the eigentime
delays were identified. The geometrical realization of those dynamical
parameters is tried in this work, which allows us to give a
geometrical derivation of the Beutler-Fano formulas appearing in
eigenphase shifts and time delays.

The geometrical realization is based on the real three-dimensional
space  spanned by the Pauli matrices $\sigma_x$, $\sigma_y$,
$\sigma_z$ as basic vectors, where vectors are orthogonal in the sense
that 
\begin{equation}
{\rm Tr} (\sigma_i \sigma_j ) = 2 \delta_{ij} .
\end{equation}
Such a kind of space is called a Liouville space\cite{Fano57}.  A 2
$\times$ 2 traceless Hermitian matrix, which is generally expressed as
$\boldsymbol{\sigma} \cdot {\bf r}$ with ${\bf r}$ real, is a vector
in this Liouville space.  The magnitude of a vector in the
three-dimensional Liouville space corresponds to the degree of
anisotropy in coupling of eigenchannels with the cause that brings
about the dynamics of the dynamic operator corresponding to the
vector. The four-dimensional Liouville space including the unit matrix
{\bf 1} as another basic vector is also used for the 2 $\times$ 2
Hermitian matrix whose trace is not zero.

Resonant scattering can be separated from the background scattering in
the ${\cal S}$ matrix for the multichannel system around an isolated
resonance as ${\cal S}$ = ${\cal S}^0 (\pi_b + e^{-2i \delta_r } \pi_a
)$ where $\pi_a$ is the projection matrix to $\psi_E^{(a)}$ which is
the only type of continua interacting with the discrete
state and $\pi_b = 1 - \pi_a$\cite{Fano65}.  When the number of open
channels are limited to two, ${\cal S}^0$, $\pi_b + e^{-2i\delta_r}
\pi_a$, and ${\cal S}$ can be expressed using Pauli's spin matrices as
\begin{eqnarray*}
{\cal S}^0 &=& e^{-i(\delta_{\Sigma}^0 {\bf 1} +
\Delta_{12}^0 \boldsymbol{\sigma} \cdot \hat{z} )} ,\\
\pi_b + e^{-2i \delta_r} \pi_a &=& 
	  e^{-i (\delta_r {\bf 1} + \delta_r
\boldsymbol{\sigma} \cdot \hat{n}_t )} , \\
{\cal S} &=& e^{-i  ( \delta_{\Sigma}{\bf 1} + \delta_a
\boldsymbol{\sigma} \cdot \hat{n}_a ) } .
\end{eqnarray*}
Phase shift matrices $\boldsymbol{\Delta}^0$, $\boldsymbol{\Delta}_r$,
and $\boldsymbol{\Delta}$ which are Hermitian may be defined for
${\cal S}^0$, $\pi_b + e^{-2i\delta_r} \pi_a$, and ${\cal S}$ as 
\begin{eqnarray*}
\boldsymbol{\Delta}^0 &=& \frac{1}{2} (\delta_{\Sigma}^0 {\bf 1} +
\Delta_{12}^0 \boldsymbol{\sigma} \cdot \hat{z} ) ,
\\
\boldsymbol{\Delta}_r &=& \frac{1}{2} (\delta_r {\bf 1} + \delta_r
\boldsymbol{\sigma} \cdot \hat{n}_t ) ,
\\
\boldsymbol{\Delta}   &=& \frac{1}{2} ( \delta_{\Sigma}{\bf 1} + \delta_a
\boldsymbol{\sigma} \cdot \hat{n}_a ) 
\end{eqnarray*}
and are  vectors in the four-dimensional Liouville space.

According to the Campbell-Baker-Hausdorff formula, the phase shift
matrix $\boldsymbol{\Delta}$ is expressed as an infinite sum of
multiple commutators of $\boldsymbol{\Delta}^0$ and
$\boldsymbol{\Delta}_r$, which is difficult to use\cite{Weiss62}. The
geometric way of representing the combining rule of
$\boldsymbol{\Delta}^0$ and $\boldsymbol{\Delta}_r$ into
$\boldsymbol{\Delta}$ provides an alternative to that.  We first note
that the isotropic part $\frac{1}{2} \delta_{\Sigma}$ of
$\boldsymbol{\Delta}$ is simply obtained from those of
$\boldsymbol{\Delta}^0$ and $\boldsymbol{\Delta}_r$ as a simple
addition $\frac{1}{2} (\delta_{\sigma}^0 + \delta_r )$ and factored
out in ${\cal S}$ = ${\cal S}^0 (\pi_b + e^{-2i \delta_r} \pi_a )$.
Then the remaining anisotropic part of $\boldsymbol{\Delta}$ is
obtained from those of $\boldsymbol{\Delta}^0$ and
$\boldsymbol{\Delta}_r$ as
\[
e^{-i \Delta_{12}^0 \boldsymbol{\sigma} \cdot \hat{z}} e^{-i \delta_r
\boldsymbol{\sigma} \cdot \hat{n}_t} = e^{-i \delta_a
\boldsymbol{\sigma} \cdot \hat{n}_a} .
\]
The above identity can be expressed 
using the rotation matrices in the Liouville
space as 
\[
R_{\hat{z}} (2 \Delta_{12}^0 ) R_{\hat{n}_t} (2 \delta_r ) =
R_{\hat{n}_a} (2 \delta_a )  .
\]
This relation leads to the construction of the spherical triangle whose
vertices are the endpoints of vectors corresponding to the anisotropic
parts of the phase shift matrices but whose lengths are limited to
unity.  The lengths of the vectors are utilized as the vertex angles
of the spherical triangle. This spherical
triangle shows the  rule of combining the 
channel-channel couplings in the background scattering with the
resonant interaction to give the avoided crossing interactions in the
curves of eigenphase shifts as functions of energy.

The time delay matrix ${\cal Q}$ basically derives from the energy
derivative of the phase shift matrix $\boldsymbol{\Delta}$ of the
${\cal S}$ matrix. The phase shift matrix $\boldsymbol{\Delta}$ is a
vector in the four-dimensional Liouville space and is $\frac{1}{2} (
\delta_{\Sigma} {\bf 1} + \delta_a \boldsymbol{\sigma} \cdot \hat{n}_a
)$ as stated above. The energy derivative of $\delta_{\Sigma}$ and
$\delta_a$ without changing the direction of the vector $\hat{n}_a$
yields the ``partial delay time matrix'' given by $\hbar ( d
\delta_{{\bf 1} \Sigma} /dE + \boldsymbol{\sigma} \cdot \hat{n}_a ) d
\delta_a /dE $. Discussion on the time delay matrix due to a change in
the direction of the vector $\hat{n}_a$ or in frame transformation is
greatly facilitated by the use of the formula $\frac{1}{2} \hbar (d
\theta_a /dE) ({\cal S}^+ \boldsymbol{\sigma} \cdot \hat{y} {\cal S} -
\boldsymbol{\sigma} \cdot \hat{y} )$. The formula shows that the time
delay due to the change in frame transformation is the interference of
two terms. The first term inside the parenthesis comes from the energy
derivative of the frame transformation from the background
eigenchannels to the ${\cal S}$ matrix eigenchannels and the second
term from the energy derivative of the frame transformation from the
${\cal S}$ matrix eigenchannels to the background eigenchannels. The
first term corresponds to a vector rotated from the $\hat{y}$ vector
by the rotation $R_{\hat{n}_a} (-2 \delta_a )$. The net time delay
resulted from the interference is thus calculated by the vector
addition in the three-dimensional Liouville space.

Going back to the spherical triangle, the laws of sines and cosines
and other laws derivable from such laws such as the laws of cotangents
holding for the spherical triangle can be translated into dynamic laws
by converting $\delta_r$ and $\theta_a$ into energies according to
$\cot \delta_r$ = $-\epsilon_r$ and $\cot \theta_a$ =
$-\epsilon_a$. Two cotangent laws
\begin{eqnarray*}
\sin \Delta_{12}^0 \cot \delta_a &=& -
\sin \theta_a \cot \theta_t + \cos \theta_a \cos \Delta_{12}^0, \\
\cot \theta_f \sin \theta_t &=& - \cos (\pi - \delta_r ) \cos \theta_t +
\sin (\pi - \delta_r ) \cot \Delta_{12}^0 , 
\end{eqnarray*} 
can be shown to correspond to two Beutler-Fano formulas for $\cot
\delta_a$ and $\cot \theta_f$
\begin{eqnarray*}
\cot \delta_a &=& - \cot \Delta_{12}^0 \frac{\epsilon_a -
q_a}{\sqrt{\epsilon_a^2 +1}} ,\\
\cot \theta_f &=& 
 -\cot \theta_t \frac{\epsilon_r -q_{\tau}} {\sqrt{\epsilon_r ^2
+1}}  ,
\end{eqnarray*}
with such conversion.
Other laws also yields interesting relations among dynamical
parameters. 

The presence of the dual triangle of the spherical triangle indicates
that we can make a one-to-one correspondence between edge angles with
vertex angles so that if there is one valid relation we can make
another valid relation by interchanging angles with its one-to-one
corresponding angles. In other words, for each edge angle, we have a
conjugate vertex angle and vice versa. This conjugation relation can
be extended to $\epsilon_r$ and $\epsilon_a$ by making use of their
relation with $\delta_r$ and $\theta_a$, respectively. The full
conjugation relations among geometrical and dynamical parameters are
listed Table \ref{table:conjug}.  The duality of the spherical
triangle thus explains the symmetry found in the dynamic relations and
provides us with a systematic approach and complete symmetric
relations.  Besides this use of trigonometric laws of the spherical
triangle, the geometric construction in the Liouville space
facilitates other useful consideration.

Note that the geometrical laws holding for the geometrical objects in
the real three-dimensional Liouville space deal only with the
intrinsic nature of the dynamic couplings independent of the
characteristics of the individual system. It derives from that the
reduced energies hide the specific characteristics of the dynamic
couplings of the individual system such as the strengths of the
dynamic couplings between the discrete state and continua, the
indirect couplings between continua via discrete states, the resonance
positions, the avoided crossing point energy.  Intrinsic nature of the
dynamic couplings is concerned with the relations among eigenchannels
for various dynamic operators, the anisotropy in the channels coupling
in the $S^0$ and $S$ scattering, the anisotropy in the channel
coupling with the discrete state.  This shows both the beauty and the
limitations of the geometrical construction in the Liouville space. In
the actual application, we have to be careful when considering the
case close to the limits in coupling strength, where abnormal
behaviors take place in actual dynamic quantities, but where no
abnormality shows up in the Liouville space.

The present theory is developed for the system of one discrete state
and two continua.  It will be desirable to extend the theory to more
than two open channels and to overlapping resonances for which the
results of Refs. \cite{Lyuboshitz77,Simonius74} will be a great help.
It might be also valuable to apply the present theory to MQDT. In
connection with the latter, it might be interesting to apply the
present theory to extend the Lu-Fano plot to multi-open channel case.

\acknowledgements  This work was supported
by KOSEF under contract No. 961-0305-050-2 and  by
Korean Ministry of 
Education through Research Fund No. 1998-015-D00186.

\appendix

\section{Separation of the resonant contribution from the background
one in $S$ matrix}
\label{App:deriv_na_nt}

The $S$ matrix for the multichannel system in the neighborhood of  an
isolated resonance given in Eq. (\ref{S_res}) can also be written as
\begin{equation}
S_{jk} = \sum_{l} S_{jl}^0 \left( \delta_{lj} + 2\pi i
\frac{V_{lE}V_{kE}^*}{E-E_0 -i \Gamma /2 } \right) ,
\label{S_res2}
\end{equation}
where $V_{lE}$ is defined in terms of the asymptotic state
$\psi_E^{-(j)}$ as $(\psi_E^{-(j)} | H | \phi )$.
Using the relation $(E-E_0 +i\Gamma /2)/(E-E_0 -i \Gamma /2)$ =
$e^{-2i \delta_r }$ , Eq. (\ref{S_res2}) becomes
\begin{equation}
S_{jk} = \sum_{l} S_{jl}^0 \left[ \delta_{lj} + \left(e^{-2i
\delta_r}-1 \right) \frac{2\pi}{\Gamma} 
V_{lE}{V_{kE}^*} \right] .
\label{S_res3}
\end{equation}
If we consider  Fano's $\psi_{E}^{(a)}$ state defined as
\begin{equation}
| \psi_E^{(a)} \rangle = \sqrt{\frac{2\pi}{\Gamma}} \sum_j |
\psi_E^{-(j)}\rangle  V_{jE}
,
\end{equation}
and the projection operator $| \psi_E^{(a)} \rangle \langle
\psi_E^{(a)} |$ into that state, and define the projection matrix
$\Pi_a$ whose  ($l,j$) element
is given by  
\begin{equation}
( \Pi_a ) _{lj} = \langle \psi_E^{-(l)} |  \psi_E^{(a)} \rangle
\langle \psi_E^{(a)}  | \psi_E^{-(j)} \rangle = \frac{2\pi}{\Gamma} 
V_{lE} V_{jE}^* ,
\end{equation}
Eq. (\ref{S_res3}) becomes in matrix form as
\begin{equation}
S = S^0 + (e^{-2i \delta_r } -1) S^0  \Pi_a  = S^0 \left(  \Pi_b  + e^{-2i
\delta_r}  \Pi_a  \right) ,
\label{S_proj}
\end{equation}
where $ \Pi_b $ is equal to $1- \Pi_a $ and is the projection matrix
into the space orthogonal to the space of $\psi_E^{(a)}$ and may be
regarded as a projection matrix into the background. Using
Eq. (\ref{S_proj}), the time delay matrix $Q$ becomes
\begin{equation}
Q = 2 \hbar \frac{d \delta_r}{dE}  \Pi_a  .
\label{QPa}
\end{equation}
Notice that Fano's $\psi_E^{(a)}$ and other continua orthogonal to it
are the eigenstates of the time delay matrix $Q$ for the multichannel
system in the neighborhood of an isolated resonance. 

Now let us confine the number of open channels of the system to
2. Then the background scattering matrix $S^0$ and projection
matrices $\Pi_a$ and $\Pi_b$ in Eq. (\ref{S_proj}) and (\ref{QPa})
can be expressible in terms of Pauli matrices.  When we convert them
into ones in terms of Pauli matrices, it may be convenient to choose
the background eigenstates as a basis. Let us denote the background
scattering matrix and projection operators in the basis of background
eigenstates as ${\cal S}^0$, $\pi_a$, and $\pi_b$. They can be
obtained from those in the basis of asymptotic channel wavefunctions
as $\tilde{U}^0 S^0 U^0$, $\tilde{U}^0 \Pi_a U^0$, $\tilde{U}^0 \Pi_b
U^0$, respectively. Eqs. (\ref{S_proj}) and (\ref{QPa}) are now in 
the basis of background
eigenchannels as follows
\begin{eqnarray}
{\cal S} &=&  {\cal S}^0 \left(  \pi_b  + e^{-2i
\delta_r}  \pi_a  \right) , 
\label{S_proj_new}
\\
{\cal Q} &=& 2 \hbar \frac{d \delta_r}{dE}  \pi_a  .
\label{QPa_new}
\end{eqnarray}
Now in the basis of the background eigenchannels
the background scattering
matrix ${\cal S}^0$ takes the form
\begin{equation}
{\cal S}^0 = \left( \begin{array} {cc} e^{-2i\delta_1^0} & 0\\ 0& e^{-2i
\delta_2^0} \end{array} \right) = e^{-i (\delta_{\Sigma}^0 {\bf 1} +
\Delta_{12}^0 \boldsymbol{\sigma} \cdot \hat{z})}  .
\label{S0}
\end{equation}
Comparing Eq. (\ref{Q_new_basis}) and Eq. (\ref{QPa_new}), we obtain 
\begin{equation}
 \pi_a  = \frac{1}{2} \left( {\bf 1} + \boldsymbol{\sigma} \cdot \hat{n}_t
\right) ,
\label{Pa}
\end{equation}
from which $ \pi_b $ is obtained as
\begin{equation}
 \pi_b  = \frac{1}{2} \left( {\bf 1} - \boldsymbol{\sigma} \cdot \hat{n}_t
\right) .
\label{Pb}
\end{equation}
Inserting Eqs. (\ref{Pa}) and (\ref{Pb}) into Eq. (\ref{S_proj_new}), we
obtain 
\begin{equation}
{\cal S} = e^{-i (\delta_{\Sigma}^0 {\bf 1} + \Delta_{12}^0
\boldsymbol{\sigma} \cdot \hat{z} ) } e^{-i( \delta_r {\bf 1} +
\delta_r \boldsymbol{\sigma} \cdot \hat{n}_t )} .
\label{S_proj_sm}
\end{equation}
Equating Eqs. (\ref{S_new_basis}) and (\ref{S_proj_sm}), Eq. (\ref{na_nt})
is obtained.

\section{Multiplication of two rotations and spherical triangle}
\label{App:mul_rot}

Multiplication of two rotations yields another rotation.  The question
is how the three rotations are related geometrically in real
three-dimensional space. If three rotations are represented by three
rotation axes $\hat{n}_A$, $\hat{n}_B$, $\hat{n}_C$
and corresponding three rotation angles $2A$, $2B$, $2\pi - 2C$, we
found that 
three rotation axis form a spherical triangle whose vertex angles are
given by $A$, $B$, and $C$ (the corresponding opposite edge angles
will be denoted as $a$, $b$, and $c$).
When this theorem is expressed
in terms of unitary matrices in the complex two dimensional space, it
takes the form
\begin{equation}
e^{-i A \boldsymbol{\sigma} \cdot \hat{n}_A} e^{-i B
\boldsymbol{\sigma} \cdot \hat{n}_B} = e^{-i (\pi - C )
\boldsymbol{\sigma} \cdot \hat{n}_C} .
\label{exp_ABC}
\end{equation}
The proof of this theorem can easily be done if we use the relation
$e^{-i a \boldsymbol{\sigma} \cdot \hat{n}}$ = $\cos a -i
\boldsymbol{\sigma} \cdot \hat{n} \sin a$ and various sine and cosine
laws of the spherical triangle and will be omitted. The similarity
transformation $e^{-i (\pi - C ) \boldsymbol{\sigma} \cdot \hat{n}_C} h
e^{i (\pi - C ) \boldsymbol{\sigma} \cdot \hat{n}_C}$ is equal to
$e^{-i B \boldsymbol{\sigma} \cdot \hat{n}_B} e^{-i A
\boldsymbol{\sigma} \cdot \hat{n}_A}h e^{i A \boldsymbol{\sigma}
\cdot \hat{n}_A} e^{i B \boldsymbol{\sigma} \cdot \hat{n}_B}$ owing
to the identity (\ref{exp_ABC}). Such an equality of two similarity
transformation can be expressed in terms of rotation matrices as
\begin{equation}
R_{\hat{n_A}} (2A) R_{\hat{n}_B} (2B) = R_{\hat{n}_C} ( -2C)  .
\label{Rsp}
\end{equation}
Sometimes, it might be convenient if rotations in the identity
(\ref{exp_ABC}) are expressed in terms of basic axes instead of
arbitrary ones.  Without loosing generality, let us take $\hat{n}_A$
as the $z$ axis and let $\hat{n}_B$ lie on the $zx$ plane. Then
the spherical polar coordinates of three vectors are $\hat{n}_A$ =
(1,0,0), $\hat{n}_B$ = (1,$c$,0), and $\hat{n}_C$ = (1,$b$,$A$). Using
$R_{\hat{n}_B} (2B)$ = $R_{\hat{y}} (c) R_{\hat{z}} (2B) R_{\hat{y}}
(-c)$ and similar formula for $R_{\hat{n}_C} (-2C)$, Eq. (\ref{Rsp})
becomes
\begin{equation}
R_{\hat{n_A}} (2A) R_{\hat{y}} (c) R_{\hat{z}} (2B) R_{\hat{y}} (-c) =
R_{\hat{z}} (A) R_{\hat{y}} (b) R_{\hat{z}} (-2C) R_{\hat{y}} (-b)
R_{\hat{z}}(-A) .
\label{Rsp2}
\end{equation}
The validity of Eq. (\ref{Rsp2})  can be proved by inserting the
explicit formula for the rotation matrices, for example,  such as
\begin{equation}
R_{\hat{z}} (\theta ) = \left( \begin{array}{ccc} 
\cos \theta & - \sin\theta& 0 \\
\sin \theta &   \cos\theta& 0 \\
0&0&1 \end{array} \right) ,
\end{equation}
and by using sine and cosine laws of spherical triangle, the
derivation of which is straightforward but requires patience.  The
equality of two relations (\ref{Rsp}) and (\ref{Rsp2}) can be shown
using the finite rotation formula for the vector ${\bf r}'$ which is
obtained by rotating ${\bf r}$ about the unit vector $\hat{n}$ through
an angle $\Phi$ as follows
\begin{equation}
{\bf r}' = \hat{n} (\hat{n} \cdot {\bf r}) + [{\bf r} - \hat{n}
(\hat{n} \cdot {\bf r})] \cos \Phi + ({\bf r} \times \hat{n} ) \sin
\Phi .
\end{equation}

\pagebreak

\begin{table}
\caption{Conjugate variables.}
\begin{tabular}{cc}
Variable& Conjugate variable  \\ \hline
$\delta_r$ & $\pi - \theta_a$ \\
$\theta_f$ & $\delta_a$ \\
$\Delta_{12}^0$ &  $\pi - \theta_t$ \\
$\epsilon_r$ &  $- \epsilon_a$ \\
$\epsilon_{{\rm BF},r}$ &  $ \epsilon_{{\rm
BF},a}$ \\
$q_{\tau}$ & $- q_a$ \\
\end{tabular}
\label{table:conjug}
\end{table}

\begin{table}
\caption{Parameters for the model intermolecular system A$\cdots$B$_2$.}
\label{potential}
 \begin{tabular}{lr}
Parameter&Value \\
\hline
\multicolumn{2}{c}{Reduced mass between A and B$_2$}\\
m&  6756.8 a.u.\\
\multicolumn{2}{c}{Morse potential parameters} \\
D$_{\rm AB}$ &0.0034 eV\\
$\alpha_{\rm AB}$ & 1.0 a.u.$^{-1}$\\
R$_{\rm AB}^{(o)}$ & 6.82 a.u.\\
D$_{\rm CM}$ &0.00195 eV\\ 
$\alpha_{\rm CM}$ & 1.0 a.u.$^{-1}$\\
R$_{\rm CM}^{(o)}$ & 6.65 a.u.\\ 
\multicolumn{2}{c} {Van der Waals potential parameters} \\
C$_{60}$ & 0.75  eV(a.u.)$^{-6}$\\ 
C$_{62}$ & 0.119 eV(a.u.)$^{-6}$\\
C$_{80}$ & 1.58  eV(a.u.)$^{-8}$\\
C$_{82}$ & 0.8   eV(a.u.)$^{-8}$\\
\end{tabular}
\end{table}

\begin{table}
\caption{Diatomic molecular parameters.}
 \label{diatom}
\begin{tabular}{lcr} 
\multicolumn{2}{c} {Parameter} & Value \\
\hline
vibrational frequency &$\omega_e$ & 0.0162 eV \\
rotational constant& $B$ & 0.01758 meV \\
equilibrium bond length&$r_e$ & 3.044 a.u. \\
reduced mass &$\mu$& 32576.6 a.u.\\ 
\end{tabular}
\end{table}

\begin{table}
\caption{Comparison of parameters obtained from the theory in this
 paper and from the data-fitting of partial delay times calculated by
 the close-coupling method. The channels included in the
 close-coupling equation are described in the text. Values of
 parameters in Tables \ref{potential} and \ref{diatom} are used in the
 calculation.  The three energy intervals used for the data-fitting in
 the close-coupling method correspond to [$-2,2$], [$-4,4$], and
 [$-6,6$], respectively, in $\epsilon_r$ scale.  In this system, $\cot
 ^2 \theta_t \gg 1$ and $\theta_t$ $\approx$ $- \pi$. This means that
 $q_{\tau} $ $\approx$ $- \cot \Delta_{12}^0$, corresponding to the
 limiting case considered in Eq. (\ref{td_theta_t_zero}). Note that
 $q_{\tau}$ lies in the second interval [$-4,-2$] beyond the first
 interval. The avoided crossing point energy $E_a$ is located at $
 -4.4\times 10^{-6}$ eV from the resonance energy $E_0$ and is very
 close to $q_{\tau}$ when measured in $\epsilon_r$ scale since it is
 equal to $q_{\tau} \cos ^2 \theta_t$. $\sin \theta_t$ $\approx$ 0
 means that $\Gamma_2$ $\approx$ 0, physically corresponding to the
 system where the decay process into one open channel is much faster
 than that into another open channel. }
\label{tab:fit3044}
\begin{tabular}{ccccc} 
&Theory&\multicolumn{3}{c}{Close-coupling method}\\ \cline{3-5}
&in this paper&[$E_{\circ}-\Gamma ,E_{\circ}+\Gamma$]&[$E_{\circ}-
2\Gamma ,E_{\circ}+2\Gamma$]&
[$E_{\circ}-3\Gamma ,E_{\circ}+3\Gamma$] \\ \hline
$E_{\circ}$(eV)&0.01396397&0.01396567&0.01396567&0.01396567\\
$\Gamma$(eV)&2.38$\times 10^{-6}$&2.35$\times 10^{-6}$&
2.34$\times 10^{-6}$&
2.320$\times 10^{-6}$\\ 
\multicolumn{5}{c}{$2 \hbar \frac{d\delta_+}{dE}$}
\\
$q_{\tau}$&-3.69&-3.56&-3.25&-3.45\\
$\cot ^2 \theta_t$&5274&5000&1.0$\times 10^{5}$&2.7$\times 10^{5}$\\
$\lambda$&&$10^{-1}$&$10^{-9}$&$10^{-4}$\\ 
\multicolumn{5}{c}{$2 \hbar \frac{d\delta_-}{dE}$}
\\
$q_{\tau}$&-3.69&-3.56&-3.40&-3.51\\
$\cot^2 \theta_t$&5274&5000&5.9$\times 10^{4}$&-1.4$\times 10^{4}$\\
$\lambda$&&$10^{-1}$&$10^{-3}$&$10^{-5}$\\ 
\end{tabular}
\end{table}

\begin{table}
\caption{Comparison of parameters obtained from the theory in this
paper and from the data-fitting of the partial delay times calculated
by the close coupling method. The channels included in the
close-coupling equation and values of parameters used in the
calculation are the same as those in Table \ref{tab:fit3044} except
that $r_e$ = 5.044a.u. Avoided crossing point energy is located at
$8.1 \times 10^{-8}$eV from the resonance energy and lies very close
to the resonance energy, meaning that $\Delta_{12}^0$ $\approx$
$\pi/2$. Then $q_{\tau} \sim 0$.}
\label{tab:fit5044}
\begin{tabular}{ccccc} 
&Theory&\multicolumn{3}{c}{Close-coupling method}\\ \cline{3-5}
&in this paper&[$E_{\circ}-\Gamma ,E_{\circ}+\Gamma$]&
[$E_{\circ}-2\Gamma ,E_{\circ}+2\Gamma$]&
[$E_{\circ}-3\Gamma ,E_{\circ}+3\Gamma$] \\ \hline
$E_{\circ}$(eV)&0.01553487&0.01553486&0.01553486&0.01553486\\
$\Gamma$(eV)&3.35$\times 10^{-6}$&3.29$\times 10^{-6}$&
3.27$\times 10^{-6}$&
3.24$\times 10^{-6}$\\ 
\multicolumn{5}{c}{$2 \hbar \frac{d\delta_+}{dE}$}
\\
$q_{\tau}$&0.115&0.164&0.199&0.254\\
$\cot^2 \theta_t$&0.73&0.71&0.75&0.78\\
$\lambda$&&$10^{-8}$&$10^{-5}$&$10^{-5}$\\ 
\multicolumn{5}{c}{$2 \hbar \frac{d\delta_-}{dE}$}
\\
$q_{\tau}$&0.115&0.109&0.132&0.133\\
$\cot^2 \theta_t$&0.73&0.71&0.74&0.77\\
$\lambda$&&$10^{-3}$&$10^{-7}$&$10^{-3}$\\ 
\end{tabular}
\end{table}

\begin{table}
\caption{Comparison between line profile indices of partial
photo-dissociation cross sections obtained by the configuration theory
and by the close-coupling method. Adopted from Ref. [29].}
\label{tab:partial_photodissociation_cross_section}
\begin{tabular}{ccccccc} 
&\multicolumn{3}{c}{Configuration interaction
method}&\multicolumn{3}{c}{Close-coupling method}\\ $j$&$\sigma_j^o$
[arb. unit] &$\Re(q_j )$&$[\Im(q_j )]^2$&$\sigma_j^o$ [arb. unit] &
$\Re(q_j )$&$[\Im(q_j )]^2$\\ \hline
0&0.221&-356&24280&0.219&-358&24920\\
2&0.279&-233&169&0.281&-233&-48\\
4&0.305&-171&23810&0.310&-173&23040\\
6&0.282&-275&10930&0.281&-275&11390\\
8&0.154&-248&34190&0.150&-248&36880\\
10&0.035&143&76130&0.034&125&81320\\
\end{tabular}
\end{table}

\pagebreak

\begin{figure}
\caption{Interference between two time delay processes due to the
change in frame transformation.}
\label{fig:frmchg}
\end{figure} 

\begin{figure}
\caption{Four coordinate systems $xyz$, $x'y'z'$, $x''y''z''$, and
$x'''y'''z'''$ pertaining to the eigenchannels of ${\cal S}^0$, ${\cal
S}$ or
$\boldsymbol{\sigma} \cdot \pmb{\cal P}_a$, $\boldsymbol{\sigma} \cdot
\pmb{\cal P}_f$, and $\boldsymbol{\sigma} \cdot \pmb{\cal P}_t$.  }
\label{fig:euler}
\end{figure}

\begin{figure}
\caption{A diagram showing the spherical polar coordinates of
$\hat{n}_t$ in the $xyz$ and $x'y'z'$ coordinate systems. The
projection of $\hat{n}_t$ on the $x'y'$ plane coincides with
$\hat{n}_f$ except for the length.  Note that $\hat{n}_a$,
$\hat{n}_t$, and $\hat{n}_f$ lie on the same plane. This diagram shows
that the spherical polar coordinates of $\hat{n}_t$ in two coordinate
systems are given by ($1,\theta_t ,-\Delta_{12}^0$) and ($1,\theta_f
,-\delta_a$), respectively.}
\label{fig:pol_vec}
\end{figure}

\begin{figure}
\caption{The spherical triangle whose vertices are the endpoints of
$\hat{z}$, $\hat{n}_a$, and $\hat{n}_t$.}
\label{fig:sph_tri}
\end{figure}

\begin{figure}
\caption{The dual spherical triangle corresponding to the spherical
triangle of
Fig. \ref{fig:sph_tri}, where the edge angles of the latter becomes
the vertex angles of the former and vice versa.}
\label{fig:dual_st}
\end{figure}

\begin{figure}
\caption{A diagram showing the geometrical traversal of the vertex P
as energy increases from $-\infty$ to $\infty$, the point P traverses
from the point A to the opposite point $-$A along the great circle.}
\label{fig:sphere}
\end{figure}

\begin{figure}
\caption{Partial delay times $2\hbar d\delta_{\pm} /d\epsilon_r$
vs. $\epsilon_r$ are plotted for three different profile indices
$q_{\tau}$ = 0.6, 1, and 5 with $\Delta_{12}^0$ = $\pi /3$. Graphs for
$2\hbar d\delta_{+} /d\epsilon_r$ are located more shifted to the left
hand side than the corresponding ones for $2\hbar
d\delta_{-}/d\epsilon_r$. The values 0.6 of $q_{\tau}$ is close to its
possible minimum value 0.577 (=$\cot\Delta_{12}^0$). }
\label{fig:dpm_df}
\end{figure}


\begin{references}
\bibitem{ChannelTerm} The terminology ``asymptotic channel
wavefunctions'' is used for $\psi_E^{-(l)}$ ($l$ = 1,2,...,n) which
satisfy the incoming wave boundary conditions. The ``background
eigenchannel wavefunctions'' are the eigenvectors $\psi_E
^{(k)}$ of the $S^0$ matrix and related to $\psi_E^{-(l)}$ by
$\psi_E^{(k)}$ = $-ie^{i\delta_k^0} \sum_l \tilde{U}_{kl}^0
\psi_E^{-(l)}$ where $U^0$ is defined as $S^0$ = $U^0 e^{-2i \delta^0}
\tilde{U}^0$.
\bibitem{Qmatrix} $Q$ matrix in terms of $S$ matrix can take various
forms depending on which boundary conditions, incoming or outgoing,
are used and how $S$ matrix is defined. The relation $Q$ = $i \hbar
S^+ dS/dE$ used in this paper corresponds to the incoming wave
boundary condition with the $S$ matrix defined by $_{\infty}\psi_j =
\Phi_j - \sum_k \Phi_k^* S_{kj}$ where $\Phi_j$ is the outgoing wave
$v_j^{-1/2} e^{ik_j x} \omega_j (y)$. Smith used the outgoing wave
boundary condition with $S_{jk}$'s multiplying $\Phi_k$ in front of it
and obtained the different relation $Q$ = $i \hbar S dS^+ /dE$. For
notations and derivation, see Ref. \cite{Smith}.
\bibitem{Convention1} Without loosing generality, the convention that
$0 \le \Delta_{12}^0 \le \pi$ and $0 \le \theta_t \le \pi$ is chosen
so that $\sin \Delta_{12}^0 \ge 0$ and $\sin \theta_t \ge 0$. The
convention $\sin \theta_t \ge 0$ ensures that $\sin \theta_f \ge 0$
and $\tau_f = \tau_r \sin \theta_f \ge 0$ since $\theta_t \le \theta_f
\le \pi - \theta_t$.
\bibitem{Convention2} The convention of $\cos \theta = -\epsilon
/\sqrt{\epsilon^2 
+1}$ and $\sin \theta = 1/\sqrt{\epsilon^2 +1}$  mentioned in Ref. 
\cite{Lee98} is employed for  relations $\cot \theta = -
\epsilon$ like $\cot \delta_r$ = $-\epsilon_r$, $\cot \theta_a$ =
$-\epsilon_a$, $\cot \delta_a$ = $-\epsilon_{{\rm BF,a}}$, $\cot
\delta_r$ = $-\epsilon_{{\rm BF,r}}$.
\bibitem{NapierRule} The dual spherical triangle in
Fig. \ref{fig:dual_st} becomes a right triangle when $\theta_t$ =
$\pi/2$, for which Napier's rule holds\cite{ChiangBook}. According to
Napier's rule, the sine of any middle part of the circular form given
by
\[ 
\left( (\theta_f -
\frac{\pi}{2})(\pi - \delta_r )(\pi - \Delta_{12}^0 )(\theta_a -
\frac{\pi}{2} )(\frac{\pi}{2} - \delta_a ) \right)
\]
is the product of the tangents of the adjacent parts.
Eq. (\ref{cot_parallel}) is obtained by choosing $\theta_a - \pi/2$ as
the middle part.
\end{references}
\end{document}